\documentclass[11pt,tightenlines,superscriptaddress,floatfixolumn,english,aps,
notitlepage,nofootinbib]{revtex4-1}

\usepackage[utf8]{inputenc}
\usepackage{amsmath}
\usepackage{framed}
\usepackage{graphicx}
\usepackage{subfigure}
\usepackage{amsfonts}
\usepackage{fancybox}
\usepackage{tabto}
\usepackage{hyperref}

\usepackage{enumitem}
\usepackage{amssymb}
\usepackage{textcomp}
\usepackage{float}

\newcommand{\mf}{\frac}
\newcommand{\ml}{\left}
\newcommand{\mr}{\right}
\newcommand{\mla}{\langle}
\newcommand{\mra}{\rangle}
\newcommand{\msg}{\sigma}
\newcommand{\mrho}{\varrho}
\newcommand{\meps}{\varepsilon}

\begin{document}

\title{Cumulants from fluctuating width of rapidity distribution}

\author{Micha{\l} Barej}
\email{michal.barej@fis.agh.edu.pl}
\affiliation{AGH University of Science and Technology,
Faculty of Physics and Applied Computer Science,
30-059 Krak\'ow, Poland}

\author{Adam Bzdak}
\email{adam.bzdak@fis.agh.edu.pl}
\affiliation{AGH University of Science and Technology,
Faculty of Physics and Applied Computer Science,
30-059 Krak\'ow, Poland}

\begin{abstract}
In relativistic heavy-ion collisions, the longitudinal fluctuations of the fireball density caused, e.g., by baryon stopping fluctuations result in event-by-event modifications of the shape of the proton rapidity density distribution. The multiparticle rapidity correlation functions due to the varying distribution width of the proton rapidity density in central Au+Au collisions at low energies are derived. The cumulant ratios are calculated and discussed in the context of the recent STAR Collaboration results. We find that the cumulant ratios for small width fluctuations seem to be universal.
\end{abstract}

\maketitle

\section{Introduction}
In relativistic heavy-ion collisions, a very hot (about $10^{12}$~K) medium, the so-called fireball, is created in a tiny volume. Its initial shape asymmetry is reflected in the measured spectra of produced particles. The azimuthal asymmetry is broadly studied using the Fourier decomposition. In this context, the Fourier coefficients are interpreted as harmonic flows, including elliptic flow, triangular flow, and others \cite{Ollitrault:1992bk,STAR:2000ekf}. However, also the long-range longitudinal correlations can be understood as a reflection of the fireball rapidity density fluctuations \cite{Bialas:2011bz}. These fluctuations result in a nontrivial rapidity correlation function \cite{Bzdak:2012tp}. This function contributes to the proton or baryon number factorial cumulants and cumulants which are potentially very promising in the search for the predicted phase transition and critical point between hadronic matter and quark-gluon plasma \cite{Stephanov:2004wx,Braun-Munzinger:2008szb,Braun-Munzinger:2015hba,Bzdak:2019pkr,Luo:2011rg,Jeon:2000wg,Asakawa:2000wh,Gazdzicki:2003bb,Gorenstein:2003hk,Koch:2005vg,Stephanov:2008qz,Cheng:2008zh,Fu:2009wy,Skokov:2010uh,Stephanov:2011pb,Herold:2016uvv,Luo:2017faz,Szymanski:2019yho,Ratti:2019tvj,Behera:2018wqk,HADES:2020wpc,STAR:2021iop}. Therefore, it is interesting to better understand the correlations related to longitudinal fluctuations.

The longitudinal fluctuations might be affected by the baryon-stopping effect. Indeed, at high energies, baryons are produced as baryon-antibaryon pairs, satisfying the baryon number conservation. At lower energies (at which the phase transition might happen), fewer pairs are created and the greater baryon density is obtained when more incoming baryons are stopped in the specific bin of rapidity. Clearly, a change in the number of stopped baryons in the midrapidity region modifies the fireball density, as well as it should be reflected in the baryon multiplicity cumulants. Therefore, the dynamics of baryon stopping is another source of fluctuations that has to be taken into account. 

The measured proton rapidity density distribution is averaged over many events. In particular, there may be event-by-event fluctuations of the width of the distribution (described by the standard deviation) even if the total number of particles remains the same. The varying numbers of baryons stopped at different rapidity bins may be qualitatively consistent with the picture of width fluctuations. 

So far, the longitudinal fluctuations have been studied using the formalism in which the single-particle rapidity density distribution is expanded into orthogonal polynomials \cite{Bzdak:2012tp,Bzdak:2015dja,Jia:2015jga,Rohrmoser:2019xis}. The coefficients of this expansion have been measured by the ATLAS Collaboration at different collision energies and different colliding systems \cite{ATLAS:2016rbh}. This topic has been also addressed by the ALICE Collaboration \cite{ALICE:2017mtc,QuishpeQuishpe:2835858}. The width fluctuations of the single-particle rapidity density distribution modify the $a_2$ coefficient in the orthogonal polynomials formalism \cite{Bzdak:2012tp}. 

In this paper, a new approach that focuses on width fluctuations is proposed. The analytic method to extract the multiparticle rapidity correlation functions from these fluctuations is derived. Then, these correlation functions are used to calculate the corresponding factorial cumulants and cumulants. Different possible characteristics of the width fluctuations are explored. In central Au+Au collisions at low energies, the proton rapidity density distribution is well described by the Gaussian function as seen, e.g., from the recent STAR data \cite{Kimelman:2023bmt}. Close to $y=0$, it can be approximated by the quadratic function. 

This paper is organized as follows. In the next section, the general method of deriving multiparticle rapidity correlation functions from the width fluctuations is presented. Then, this method is applied to the quadratic single-particle rapidity density distribution with different width probability distributions. In the subsequent section, the examples of cumulant ratios are calculated and discussed in the context of the corresponding STAR measurements. We also argue how a different rapidity range affects the results. Finally, the comments and summary are presented. Higher-order cumulants are discussed in the appendixes.

\section{Factorial cumulants from width fluctuations}
Let $\mrho(y)$ be a single-particle rapidity density distribution. Suppose the width of this distribution described by the standard deviation, $\msg$, fluctuates from event to event. The measured distribution $\mrho_\text{meas}(y)$ is $\mrho(y)$ averaged over $\msg$. Similarly, $\mrho_\text{meas,2}(y_1,y_2)$ is a two-particle rapidity density distribution averaged over $\msg$. Then, we construct a two-particle rapidity correlation function originating from the width fluctuation. Following the same reasoning, we calculate higher-order multiparticle rapidity correlation functions.

Suppose $\mrho(y)$ is given by the normal distribution,
\begin{equation} \label{eq:gauss-rho}
\mrho(y) = \mf{N_t}{\sqrt{2 \pi} \msg} \exp \ml(- \mf{y^2}{2\msg^2} \mr)\,,
\end{equation}
where $N_t = \int_{-\infty}^{+\infty} dy\, \mrho(y)$ is the total number of particles. This function is valid for proton rapidity density distribution in central low-energy collisions \cite{Kimelman:2023bmt}. For $y \approx 0$ (the midrapidity region), this distribution can be approximated by the quadratic function:
\begin{equation} \label{eq:quadratic-rho}
\mrho(y) \approx \mf{N_t}{\sqrt{2 \pi} \msg} \ml(1- \mf{y^2}{2\msg^2} \mr)\,.
\end{equation}

\begin{figure}[H]
\begin{center}
    \includegraphics[width=0.5\textwidth]{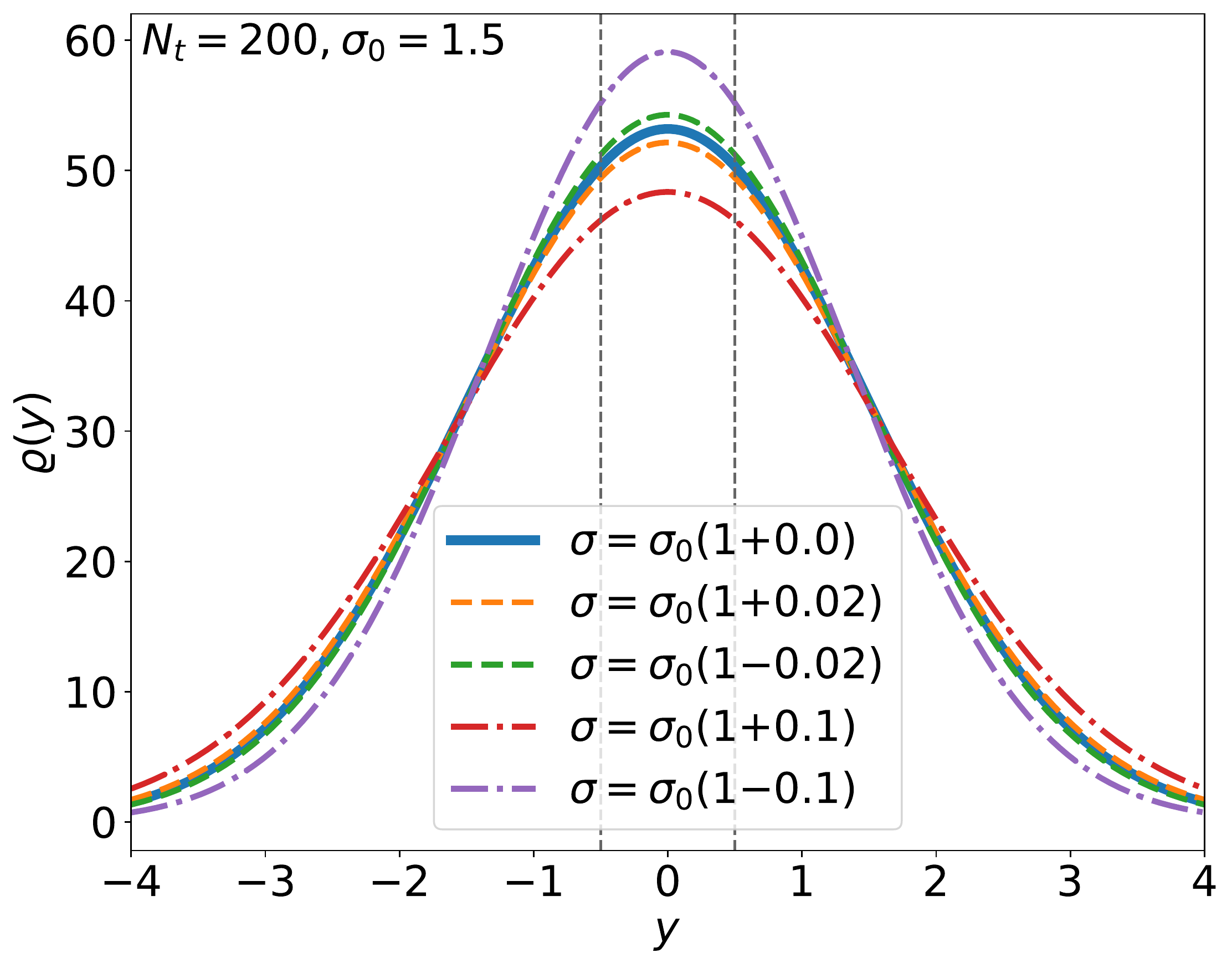}
\caption{Proton rapidity density distribution \eqref{eq:gauss-rho} with different values of $\msg$. $N_t$ and $\msg_0$ values correspond to proton distribution from central Au+Au collisions at 7.7~GeV. The vertical dashed lines show the calculation (measurement) range $y \in [-0.5, 0.5]$.}
\label{fig:sigma-modifies}
\end{center}
\end{figure}

We assume that $\sigma$ (representing the width of the single-particle rapidity density distribution) fluctuates from event to event, being a random variable following the probability distribution, $p(\sigma)$, where $\int_{0}^{+\infty} d\msg \: p(\msg) = 1$. The average value of $\sigma$ will be denoted by $\msg_0$ 
\begin{equation}
\msg_0 \equiv \mla \msg \mra = \int_{0}^{+\infty} d\msg \: \msg p(\msg) \,.
\end{equation}
To emphasize the variability of the $\msg$ parameter, we denote the rapidity distribution (\eqref{eq:gauss-rho} and \eqref{eq:quadratic-rho}) as $\mrho(y, \msg)$. 
In Fig. \ref{fig:sigma-modifies}, we show an example of how a change of $\msg$ modifies the rapidity distribution, using the values of the parameters that are later used in our examples. Clearly, an increase of $\msg$ makes the distribution wider, whereas a decrease of $\msg$, makes the distribution narrower. As seen from the figure, the chosen interval $y \in [-0.5, 0.5]$ is so narrow that $\mrho(y, \msg)$ with different $\msg$ look like being rescaled.

The measured (averaged over $\sigma$) single-particle rapidity density distribution, $\mrho_{\text{meas}}(y)$, is obtained as
\begin{equation} \label{eq:avg-dn-dy}
\mrho_{\text{meas}}(y) = \int_{0}^{+\infty} d\msg \mrho(y, \msg) p(\msg) \,.
\end{equation}
In the limiting case of $p(\msg) = \delta(\msg - \msg_0)$, where $\delta$ is the Dirac delta function, $\mrho_{\text{meas}}(y)$ becomes $\mrho(y; \msg = \msg_0)$. We note that the total number of particles, say baryons, $N_t = \int_{-\infty}^{+\infty} dy\, \mrho_\text{meas}(y)$ is unchanged.

Similarly, the averaged two-particle rapidity density distribution reads
\begin{equation}
\mrho_{\text{meas},2}(y_1, y_2) = \int_{0}^{+\infty} d\msg \mrho(y_1, \msg) \mrho(y_2, \msg) p(\msg) \,.
\end{equation}
Then, the two-particle rapidity correlation function is defined as
\begin{equation}
C_2(y_1, y_2) = \mrho_{\text{meas},2}(y_1, y_2) - \mrho_\text{meas}(y_1) \mrho_\text{meas}(y_2)  \,.
\end{equation}
The second factorial cumulant is obtained by integrating the correlation function,
\begin{equation}
\hat{C}_2 = \int_{-Y}^{Y} dy_1 \int_{-Y}^{Y} dy_2 \: C_2(y_1, y_2)\,,
\end{equation}
where $Y$ characterizes the rapidity range of the measured correlations.

By analogy, the $n$-particle rapidity density distribution is
\begin{equation} \label{eq:n-part-rap-distrib}
\mrho_{\text{meas},n}(y_1, y_2, ..., y_n) = \int_{0}^{+\infty} d\msg \mrho(y_1, \msg) \mrho(y_2, \msg) \cdots \mrho(y_n, \msg) p(\msg) \,.
\end{equation}
For example, the three-particle correlation function reads:
\begin{equation}
\begin{split}
C_3(y_1, y_2, y_3) =& \mrho_{\text{meas},3}(y_1, y_2, y_3) - \mrho_\text{meas}(y_1) \mrho_\text{meas}(y_2) \mrho_\text{meas}(y_3) - \mrho_\text{meas}(y_1) C_2(y_2, y_3)\\ 
&- \mrho_\text{meas}(y_2) C_2(y_1, y_3) - \mrho_\text{meas}(y_3) C_2(y_1, y_2)\,,
\end{split}
\end{equation}
and $n$-particle correlation functions ($n=4,5,6$) are defined, e.g., in Ref. \cite{Bzdak:2015dja}. The $n$th factorial cumulant is obtained by
\begin{equation}
\hat{C}_n = \int_{-Y}^{Y} dy_1 \cdots \int_{-Y}^{Y} dy_n \: C_n(y_1, y_2, ..., y_n)\,.
\end{equation}

\subsection{Correlation functions and factorial cumulants for the quadratic rapidity distribution.}
In our calculations, we assume the quadratic single-particle rapidity density distribution \eqref{eq:quadratic-rho} which is a good approximation in the midrapidity region. It is convenient to introduce the notation:
\begin{equation} \label{eq:m-k}
m_k = \ml\mla \mf{1}{\msg^k} \mr\mra = \int_{0}^{+\infty} d\msg \: \mf{p(\msg)}{\msg^k} \,.
\end{equation}

The $\sigma$ averaged rapidity distribution \eqref{eq:avg-dn-dy} reads
\begin{equation} \label{eq:dn-dy-quadr-avg-sig}
\mrho_\text{meas}(y) = \mf{N_t}{\sqrt{2\pi}} \ml( m_1 - \mf{1}{2} m_3 y^2 \mr) \,.
\end{equation}
The first factorial cumulant (equal to the mean number of particles in the integration interval) is given by
\begin{equation} \label{eq:c1-quadr}
\hat{C}_1 = \mla N \mra = \int_{-Y}^{Y} dy \: \mrho_\text{meas}(y) = \mf{N_t Y}{\sqrt{2\pi}} \ml( 2 m_1 - \mf{1}{3} m_3 Y^2 \mr) \,.
\end{equation}

The two-particle rapidity density distribution, two-particle correlation function, and the second factorial cumulant read
\begin{equation} \label{eq:rho2}
\mrho_{\text{meas},2}(y_1, y_2) = \ml(\mf{N_t}{\sqrt{2 \pi}}\mr)^2 \ml[m_2 - \mf{1}{2} m_4 (y_1^2 + y_2^2) + \mf{1}{4} m_6 y_1^2 y_2^2 \mr] \,,
\end{equation}
\begin{equation} \label{eq:corr2-quadr}
C_2(y_1, y_2) = \ml(\mf{N_t}{\sqrt{2 \pi}}\mr)^2 \mf{1}{2^2} \ml[4 A_0 - 2 A_1 (y_1^2 + y_2^2) + A_2 y_1^2 y_2^2 \mr] \,,
\end{equation}
\begin{equation} \label{eq:c2-quadr}
\hat{C}_2 = \ml(\mf{N_t Y}{\sqrt{2 \pi}}\mr)^2 \ml[4 A_0 - \mf{4}{3} A_1 Y^2 + \mf{1}{9} A_2 Y^4 \mr] \,,
\end{equation}
where
\begin{equation} \label{eq:c2-params}
\begin{split}
A_0 &= m_2 - m_1^2 \,, \\
A_1 &= m_4 - m_1 m_3 \,, \\
A_2 &= m_6 - m_3^2 \,.
\end{split}
\end{equation}

The three-particle rapidity density distribution, correlation function, and the third factorial cumulant read
\begin{equation} \label{eq:rho3}
\mrho_{\text{meas},3}(y_1, y_2, y_3) = \ml(\mf{N_t}{\sqrt{2 \pi}}\mr)^3 \ml[m_3 - \mf{1}{2} m_5 (y_1^2 + y_2^2 + y_3^2) + \mf{1}{4} m_7 (y_1^2 y_2^2 + y_1^2 y_3^2 + y_2^2 y_3^2) - \mf{1}{8} m_9 y_1^2 y_2^2 y_3^2 \mr] \,,
\end{equation}
\begin{equation} \label{eq:corr3-quadr}
C_3(y_1, y_2, y_3) = \ml(\mf{N_t}{\sqrt{2 \pi}}\mr)^3 \mf{1}{2^3} \ml[8A_0 - 4 A_1 (y_1^2 + y_2^2 + y_3^2) + 2 A_2 (y_1^2 y_2^2 + y_1^2 y_3^2 + y_2^2 y_3^2) - A_3 y_1^2 y_2^2 y_3^2 \mr] \,,
\end{equation}
\begin{equation} \label{eq:c3-quadr}
\hat{C}_3 = \ml(\mf{N_t Y}{\sqrt{2 \pi}}\mr)^3 \ml[8 A_0 - 4 A_1 Y^2 + \mf{2}{3} A_2 Y^4 - \mf{1}{27} A_3 Y^6 \mr] \,,
\end{equation}
where
\begin{equation} \label{eq:c3-params}
\begin{split}
A_0 &= 2 m_1^3 - 3 m_1 m_2 + m_3 \,, \\
A_1 &= 2 m_1^2 m_3 - m_2 m_3 - 2 m_1 m_4 + m_5 \,, \\
A_2 &= 2 m_1 m_3^2 - 2 m_3 m_4 - m_1 m_6 + m_7 \,, \\
A_3 &= 2 m_3^3 - 3 m_3 m_6 + m_9\,.
\end{split}
\end{equation}

The four-particle rapidity density distribution is given by
\begin{equation} \label{eq:rho4}
\begin{split}
\mrho_{\text{meas},4}(y_1, y_2, y_3, y_4) = \ml(\mf{N_t}{\sqrt{2 \pi}}\mr)^4 &\ml[ m_4 -\mf{1}{2}m_6 \sum_{i=1}^{4} y_i^2 + \mf{1}{4} m_8 \sum_{j>i} y_i^2 y_j^2 - \mf{1}{8} m_{10} \sum_{k>j>i} y_i^2 y_j^2 y_k^2 \mr. \\
&\left.+ \mf{1}{16} m_{12} y_1^2 y_2^2 y_3^2 y_4^2 \mr] \,.
\end{split}
\end{equation}
The four-particle correlation function and the fourth factorial cumulant read
\begin{equation} \label{eq:corr4-quadr}
\begin{split} 
C_4(y_1, y_2, y_3, y_4) = \ml(\mf{N_t}{\sqrt{2 \pi}}\mr)^4 \mf{1}{2^4} &\ml[-16 A_0 +8 A_1 \sum_{i=1}^{4} y_i^2 -4 A_2 \sum_{j>i} y_i^2 y_j^2 - 2 A_3 \sum_{k>j>i} y_i^2 y_j^2 y_k^2 + A_4 y_1^2 y_2^2 y_3^2 y_4^2 \mr] \,,
\end{split}
\end{equation}
\begin{equation} \label{eq:c4-quadr}
\hat{C}_4 = \ml(\mf{N_t Y}{\sqrt{2 \pi}}\mr)^4 \ml[-16 A_0 + \mf{32}{3} A_1 Y^2 - \mf{8}{3} A_2 Y^4 - \mf{8}{27} A_3 Y^6 + \mf{1}{81} A_4 Y^8 \mr] \,,
\end{equation}
where
\begin{equation} \label{eq:c4-params}
\begin{split}
A_0 &= 6 m_1^4 - 12 m_1^2 m_2 + 3 m_2^2 + 4 m_1 m_3 - m_4 \,, \\
A_1 &= 6 m_1^3 m_3 - 6 m_1 m_2 m_3 + m_3^2 - 6 m_1^2 m_4 + 3 m_2 m_4 + 3 m_1 m_5 - m_6 \,, \\
A_2 &= 6 m_1^2 m_3^2 - 2 m_2 m_3^2 - 8 m_1 m_3 m_4 + 2 m_4^2 + 2 m_3 m_5 - 2 m_1^2 m_6 + m_2 m_6 + 2 m_1 m_7 - m_8 \,, \\
A_3 &= m_{10} - 6 m_1 m_3^3 + 6 m_3^2 m_4 + 6 m_1 m_3 m_6 - 3 m_4 m_6 - 3 m_3 m_7 - m_1 m_9\,, \\
A_4 &= m_{12} - 6 m_3^4 + 12 m_3^2 m_6 - 3 m_6^2 - 4 m_3 m_9\,.
\end{split}
\end{equation}

Five- and six-particle correlation functions and the corresponding factorial cumulants are presented in Appendix \ref{app:c5-c6}.

We note that $\hat{C}_k$ is proportional to $N_t^k$ which indicates the long-range correlations. This is understandable since the change of the width of the distribution modifies the distribution in the whole rapidity range.

\subsection{$\msg$ fluctuations}
\begin{figure}[h]
\begin{center}
    \includegraphics[width=0.5\textwidth]{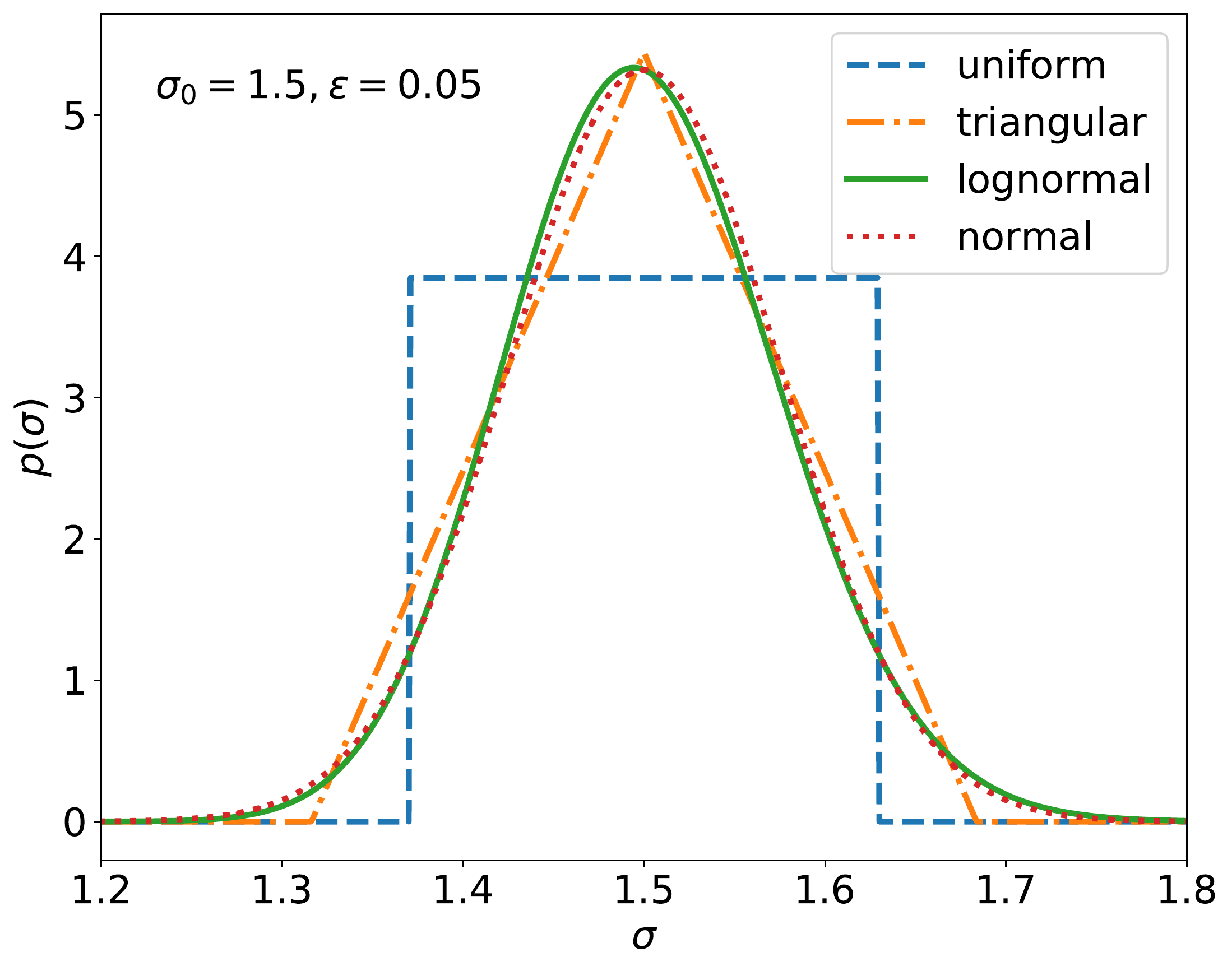}
\caption{Different $\msg$ distributions studied in this paper. They all have the same expectation $\msg_0$ and the standard deviation $\meps \msg_0$.}
\label{fig:distribs}
\end{center}
\end{figure}

We do not know how $\msg$ fluctuates in realistic heavy-ion collisions however we may study various probability distributions, $p(\msg)$, to obtain a better insight. In all the presented distributions we require that 
\begin{equation} \label{eq:expect}
\mla \msg \mra = \msg_0
\end{equation}
and the standard deviation 
\begin{equation} \label{eq:std-constr}
\sqrt{\mla(\msg - \mla \msg \mra)^2 \mra } = \meps \msg_0\,,
\end{equation}
where $\meps$ can vary and determines the strength of the $\msg$ fluctuations. For $\meps=0$, there are no $\msg$ fluctuations and all the correlations vanish. Various $\msg$ distributions with $\msg_0$, and $\meps$ used in our calculations are presented in Fig. \ref{fig:distribs}.

\subsubsection{Uniform distribution} \label{sec:unif}
First, we assume that $\msg$ follows a uniform distribution. It is the simplest distribution to study however most likely it is not very realistic. Namely,
\begin{equation} \label{eq:unif}
p(\msg)  = \begin{cases} \mf{1}{2 \sqrt{3} \meps \msg_0} &\text{if } \msg \in [\msg_0 (1 - \sqrt{3} \meps), \msg_0 (1 + \sqrt{3} \meps)], \\
0 & \text{otherwise }\,,
\end{cases}
\end{equation}
where we assume $0 \leq \meps \leq \mf{1}{\sqrt{3}}$. $\sqrt{3}$ appears because of the required standard deviation \eqref{eq:std-constr}. We obtain\footnote{We note that $m_1 = \lim\limits_{k \to 1} m_k$.},
\begin{equation} \label{eq:mk-unif}
\begin{split}
m_1 &= \mf{1}{2 \sqrt{3} \meps \msg_0} \ln \ml( \mf{1 + \sqrt{3}\meps}{1 - \sqrt{3}\meps} \mr)\,,\\
m_k &= \mf{1}{2\sqrt{3}\meps \msg_0^k} \mf{(1- \sqrt{3}\meps)^{1-k} - (1 + \sqrt{3}\meps)^{1-k}}{k-1} \,, \quad k = 2, 3, ...
\end{split}
\end{equation}
One can calculate full analytic expressions for the factorial cumulants by applying Eq. \eqref{eq:mk-unif} to Eqs. \eqref{eq:c1-quadr}, \eqref{eq:c2-quadr}, \eqref{eq:c3-quadr}, and \eqref{eq:c4-quadr}. $\meps$ is expected to be small, so we can expand full expressions into power series in terms of $\meps$ about 0. Here we present the leading-order terms (lowest power in $\meps$):
\begin{equation} \label{eq:c1-unif-approx}
\hat{C}_1 \approx \mf{N_t z}{\sqrt{2 \pi}} \ml(2 - \mf{1}{3} z^2 \mr)\,,
\end{equation}
\begin{equation} \label{eq:c2-unif-approx}
\hat{C}_2 \approx \ml(\mf{N_t z}{\sqrt{2 \pi}}\mr)^2 \meps^2 \ml(2 - z^2 \mr)^2\,,
\end{equation}
\begin{equation} \label{eq:c3-unif-approx}
\hat{C}_3 \approx \ml(\mf{N_t z}{\sqrt{2 \pi}}\mr)^3 \mf{24 \meps^4}{5} \ml(2 - z^2  \mr)^2 \ml(1 - z^2 \mr)\,,
\end{equation}
\begin{equation} \label{eq:c4-unif-approx}
\hat{C}_4 \approx -\ml(\mf{N_t z}{\sqrt{2 \pi}}\mr)^4 \mf{6 \meps^4}{5} \ml(2 - z^2  \mr)^4\,,
\end{equation}
where
\begin{equation} \label{eq:def-z}
z = \mf{Y}{\msg_0} \,.
\end{equation}
The factorial cumulants $\hat{C}_5$ and $\hat{C}_6$ are presented in Appendix \ref{app:c5-c6}.

\subsubsection{Triangular distribution} \label{sec:trian}
A step towards a more realistic $\msg$ distribution is the triangular distribution. Namely,
\begin{equation} \label{eq:trian}
p(\msg)  = \begin{cases} \mf{\msg - \msg_0(1 - \sqrt{6}\meps)}{6\meps^2 \msg_0^2} &\text{if } \msg \in [\msg_0 (1 - \sqrt{6}\meps), \msg_0], \\
\mf{-\msg + \msg_0(1 + \sqrt{6}\meps)}{6\meps^2 \msg_0^2} &\text{if } \msg \in (\msg_0, \msg_0 (1 + \sqrt{6}\meps)], \\
0 & \text{otherwise }\,,
\end{cases}
\end{equation}
where $0 \leq \meps \leq \mf{1}{\sqrt{6}}$, and $\sqrt{6}$ comes from the standard deviation constraint \eqref{eq:std-constr}.

Then\footnote{We note that $m_1 = \lim\limits_{k \to 1} m_k$ and $m_2 = \lim\limits_{k \to 2} m_k$.},
\begin{equation} \label{eq:mk-trian}
\begin{split}
m_1 &= \mf{1}{6\meps^2 \msg_0} \ml[ (1 + \sqrt{6}\meps) \ln \ml(1 + \sqrt{6}\meps \mr) + (1 - \sqrt{6}\meps) \ln \ml(1 - \sqrt{6}\meps\mr) \mr]\,,\\
m_2 &= -\mf{1}{6\meps^2 \msg_0^2} \ln \ml( 1- 6\meps^2 \mr)\,,\\
m_k &= \mf{1}{6\meps^2 \msg_0^k (k-1)(k-2) } \ml[ \mf{1}{(1-\sqrt{6}\meps)^{k-2}} + \mf{1}{(1+\sqrt{6}\meps)^{k-2}} - 2 \mr] \,, \quad k = 3, 4, 5, ...
\end{split}
\end{equation}

One can calculate full analytic expressions for $\hat{C}_k$ using Eqs. \eqref{eq:c1-quadr}, \eqref{eq:c2-quadr}, \eqref{eq:c3-quadr}, and \eqref{eq:c4-quadr}. The leading-order terms (lowest power in $\meps$) are given by
\begin{equation} \label{eq:c1-trian-approx}
\hat{C}_1 \approx \mf{N_t z}{\sqrt{2 \pi}}  \ml(2 - \mf{1}{3} z^2 \mr)\,,
\end{equation}
\begin{equation} \label{eq:c2-trian-approx}
\hat{C}_2 \approx \ml(\mf{N_t z}{\sqrt{2 \pi}}\mr)^2 \meps^2 \ml(2 - z^2 \mr)^2\,,
\end{equation}
\begin{equation} \label{eq:c3-trian-approx}
\hat{C}_3 \approx \ml(\mf{N_t z}{\sqrt{2 \pi}}\mr)^3 \mf{42 \meps^4}{5} \ml(2 - z^2  \mr)^2 \ml(1 - z^2 \mr)\,,
\end{equation}
\begin{equation} \label{eq:c4-trian-approx}
\hat{C}_4 \approx -\ml(\mf{N_t z}{\sqrt{2 \pi}}\mr)^4 \mf{3\meps^4}{5} \ml(2 - z^2  \mr)^4\,,
\end{equation}
where $z$ is given by Eq. \eqref{eq:def-z}. The higher-order factorial cumulants are presented in Appendix \ref{app:c5-c6}.

\subsubsection{Lognormal distribution} \label{sec:logn}
It would be natural to assume that $\msg$ follows the normal distribution however, by definition $\msg \geq 0$ whereas the normal distribution allows also for negative values. Also, the integrals used to calculate $m_k$, Eq. \eqref{eq:m-k}, do not converge for the normal distribution. To overcome these issues and still obtain analytic results, we assume that $\msg$ follows the lognormal distribution (its domain by definition is $(0, +\infty)$). Namely,
\begin{equation} \label{eq:lognorm}
p(\msg) = \mf{1}{\sqrt{2 \pi} b \msg} \exp\ml( - \mf{(\ln \msg - a)^2}{2b^2} \mr) \,.
\end{equation}
Its expectation $\mla \msg \mra = \exp\ml(a + \mf{b^2}{2}\mr) $ and its variance $Var(\msg) = [\exp(b^2) - 1] \exp(2 a + b^2)$. This with the constraints \eqref{eq:expect} and \eqref{eq:std-constr} gives
\begin{equation}
\begin{split}
a &= \ln\ml(\mf{\msg_0}{\sqrt{\meps^2 + 1}} \mr)\,,\\
b &= \sqrt{\ln(\meps^2 + 1)}\,.
\end{split}
\end{equation}
We have checked that such a distribution is very close to the corresponding normal distribution, as can be seen in Fig. \ref{fig:distribs}. 

For this distribution,
\begin{equation} \label{eq:mk-logn}
m_k = \mf{(1 + \meps^2)^{\mf{k(k+1)}{2} } }{\msg_0^k} \quad \text{for } k=1,2,3,...
\end{equation}

Again, one can easily calculate the full analytic expressions for $\hat{C}_k$ using Eqs. \eqref{eq:c1-quadr}, \eqref{eq:c2-quadr}, \eqref{eq:c3-quadr}, and \eqref{eq:c4-quadr}. For small $\meps$, the leading order terms read:
\begin{equation} \label{eq:c1-logn-approx}
\hat{C}_1 \approx \mf{N_t z}{\sqrt{2 \pi}} \ml(2 - \mf{1}{3} z^2 \mr)\,,
\end{equation}
\begin{equation} \label{eq:c2-logn-approx}
\hat{C}_2 \approx \ml( \mf{N_t z}{\sqrt{2 \pi}} \mr)^2 \meps^2 \ml(2 - z^2 \mr)^2\,,
\end{equation}
\begin{equation} \label{eq:c3-logn-approx}
\hat{C}_3 \approx \ml( \mf{N_t z}{\sqrt{2 \pi}} \mr)^3 3\meps^4 \ml(2 - z^2 \mr)^2 \ml(2 - 3z^2 \mr)\,,
\end{equation}
\begin{equation} \label{eq:c4-logn-approx}
\hat{C}_4 \approx \ml( \mf{N_t z}{\sqrt{2 \pi}} \mr)^4 16\meps^6 \ml(2 - z^2 \mr)^2 \ml(9 z^4 - 14z^2 + 4 \mr)\,,
\end{equation}
where $z$ is given by Eq. \eqref{eq:def-z}. The higher-order factorial cumulants are presented in Appendix \ref{app:c5-c6}. 

We note that for all three studied distributions, the factorial cumulants depend on $z = Y/ \msg_0$ and not on $Y$ and $\msg_0$ separately. The leading-order term of $\hat{C}_k$ is always proportional to $(N_t z)^k$. As expected, when $\meps \to 0$, all $\hat{C}_k=0$ ($k \geq 2$) because in this case there are no width fluctuations. We also note that the leading-order terms of $\hat{C}_1$ and $\hat{C}_2$ are universal for the three studied $p(\msg)$ distributions, whereas the higher-order terms and exact results differ among the distributions.

\subsubsection{Truncated normal distribution - numerical computations} \label{gauss}
Here we assume that $\msg$ follows the normal distribution given by
\begin{equation} \label{eq:sigma-normal}
p(\msg) = \mf{1}{\sqrt{2 \pi} \meps \msg_0 } \exp\ml( - \mf{(\msg - \msg_0)^2}{2(\meps \msg_0)^2} \mr) \,,
\end{equation}
where, as in uniform, triangular, and lognormal distributions, we require the expectation to be $\msg_0$ and the standard deviation $\meps \msg_0$. In order to require $\msg > 0$ we used the truncated normal distribution.\footnote{We used the truncated, normalized normal distribution, $\widetilde{p}(\msg) = A p(\msg)$ if $\msg \in [a, b]$ and $\widetilde{p}(\msg) = 0$ otherwise, where $p(\msg)$ is given by Eq. \eqref{eq:sigma-normal} and $A = 1 / \int_a^b d\msg p(\msg)$ with $a$ and $b$ being the limits of the $\msg$ interval.} As mentioned earlier, we were unable to obtain analytic results for the normal distribution however we calculated $m_k$ numerically using \cite{scipy-integrate}. To test this approach, we used the same numerical method for uniform, triangular, and lognormal distributions and reproduced the exact results with great precision.

\section{Cumulant ratios}
Here we show the cumulant ratios extracted from the factorial cumulants obtained for the quadratic rapidity density distribution \eqref{eq:quadratic-rho}. We use the relations between the cumulants and factorial cumulants \cite{Friman:2022wuc}.\footnote{See also Appendix A of Ref. \cite{Bzdak:2019pkr} for explicit formulas for the first six cumulants.} 

Our assumption of the Gaussian (approximately quadratic) proton rapidity density distribution is applicable for low collision energies where also the interesting anomalies of the scaled kurtosis were measured by the STAR Collaboration \cite{STAR:2021fge}. We have used the simulated net-proton rapidity density distribution in central 0-5\% Au+Au collisions at $\sqrt{s_{_{NN}}}=7.7$~GeV from Ref \cite{Vovchenko:2021kxx} to estimate that $N_t \approx 200$ and $\msg_0 \approx 1.5$ are reasonable values.\footnote{At low energies, there are very few produced antiprotons. Therefore, the net-proton rapidity density distribution is a good approximation of the proton distribution.} For $\sqrt{s_{_{NN}}} = 3\; \text{GeV}$, we have used the preliminary STAR data on proton density distribution in central 0-5\% Au+Au collisions \cite{Kimelman:2023bmt} and extracted the parameters $N_t \approx 175$, $\msg_0 \approx 0.75$. We note that the estimated $\msg_0(3~\text{GeV})/\msg_0(7.7~\text{GeV}) = 0.75/1.5 = 0.5$ is very close to the rough estimate $\ln(3) / \ln(7.7) \approx 0.54$. We choose $Y = 0.5$ as in the STAR measurements. 

In our calculations we have tried two methods of approximating the cumulant ratios. In the first method, we calculate the cumulants from the factorial cumulants using only the earlier presented leading-order terms of $\hat{C}_n$. This approximation works very well as seen in the following examples. The results of this method are denoted in the figures as ``approx with $\hat{C}$''.

In the second method, we calculate the cumulants from the factorial cumulants using exact expressions for $\hat{C}_n$ and then we approximate the obtained cumulant ratios by the power series expansion about $\meps = 0$. We obtain the expansions of the form:
\begin{equation} \label{eq:cum-ratio-expansion}
\mf{\kappa_n}{\kappa_2} = 1 + a_n f(N_t, z) \meps^2 + \underbrace{...}_{O(\meps^4)}\,,
\end{equation}
where $a_n = 6, 18, 42, 90$ for $n=3, 4, 5, 6$, respectively, and 
\begin{equation}
f(N_t,z) = \mf{N_t z}{\sqrt{2 \pi}} \mf{(2 - z^2)^2}{6-z^2}
\end{equation}
is a universal function for the studied $\msg$ distributions (uniform, triangular, lognormal) whereas higher-order terms (e.g., $\meps^4$) differ between the distributions.  We show the result of this method as ``approx with $\kappa$ up to $\meps^2$''. As seen from the following examples, this approximation works well for small $\meps$. Therefore, for small $\msg$ fluctuations, the cumulant ratio $\kappa_n/\kappa_2$ is described by the common formula for very different $\msg$ distributions. This suggests that the cumulant ratios are independent of $p(\msg)$ for small $\meps$. $f(N_t,z)$ is found to be equaivalent to considering only the leading-order terms of $\hat{C}_1$ and $\hat{C}_2$ and neglecting higher-order factorial cumulants. This indicates that for small $\meps$ the cumulant ratios are dominated by two-particle correlations. We checked that including $\meps^4$ terms improves the results but works worse than the ``approx with $\hat{C}$'' method.

\subsection{$\sqrt{s_{_{NN}}} = 7.7$~GeV}
We assume $N_t=200$, $\msg_0=1.5$, $Y=0.5$ which correspond to the STAR measurements in central Au+Au collisions at $\sqrt{s_{NN}}=7.7$~GeV.

\begin{figure}[H]
\begin{center}
    \includegraphics[width=0.46\textwidth]{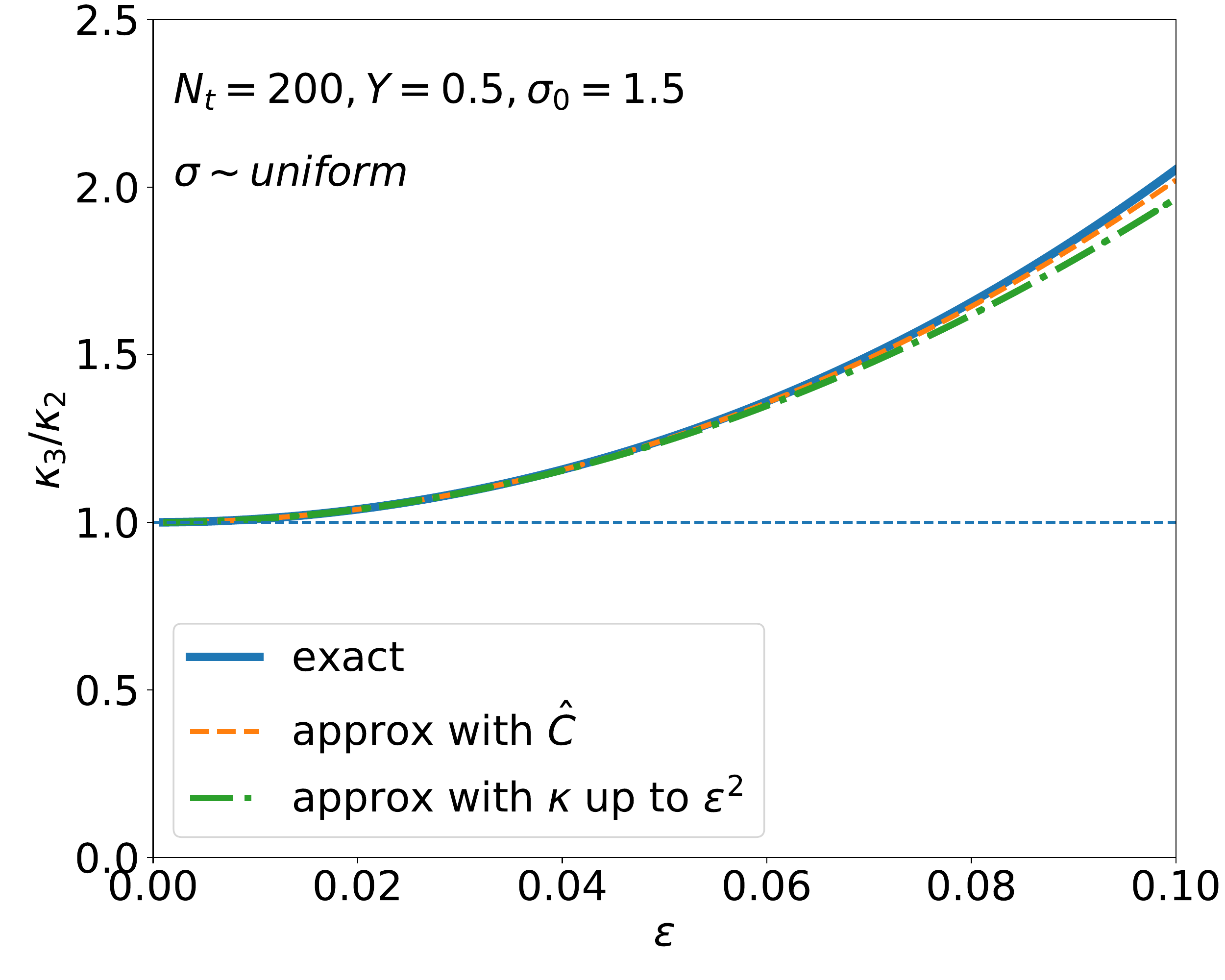}
    \includegraphics[width=0.46\textwidth]{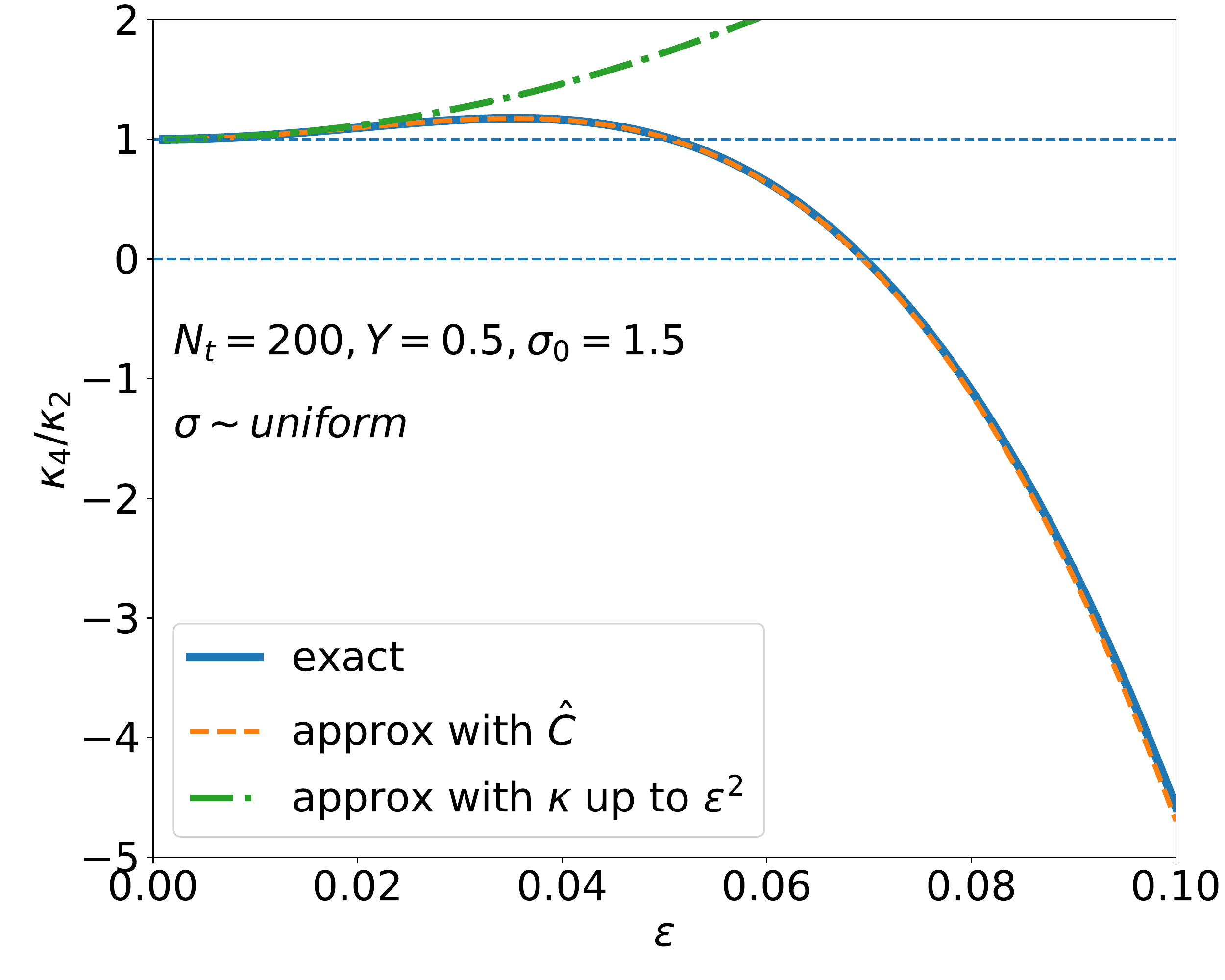}
    \includegraphics[width=0.46\textwidth]{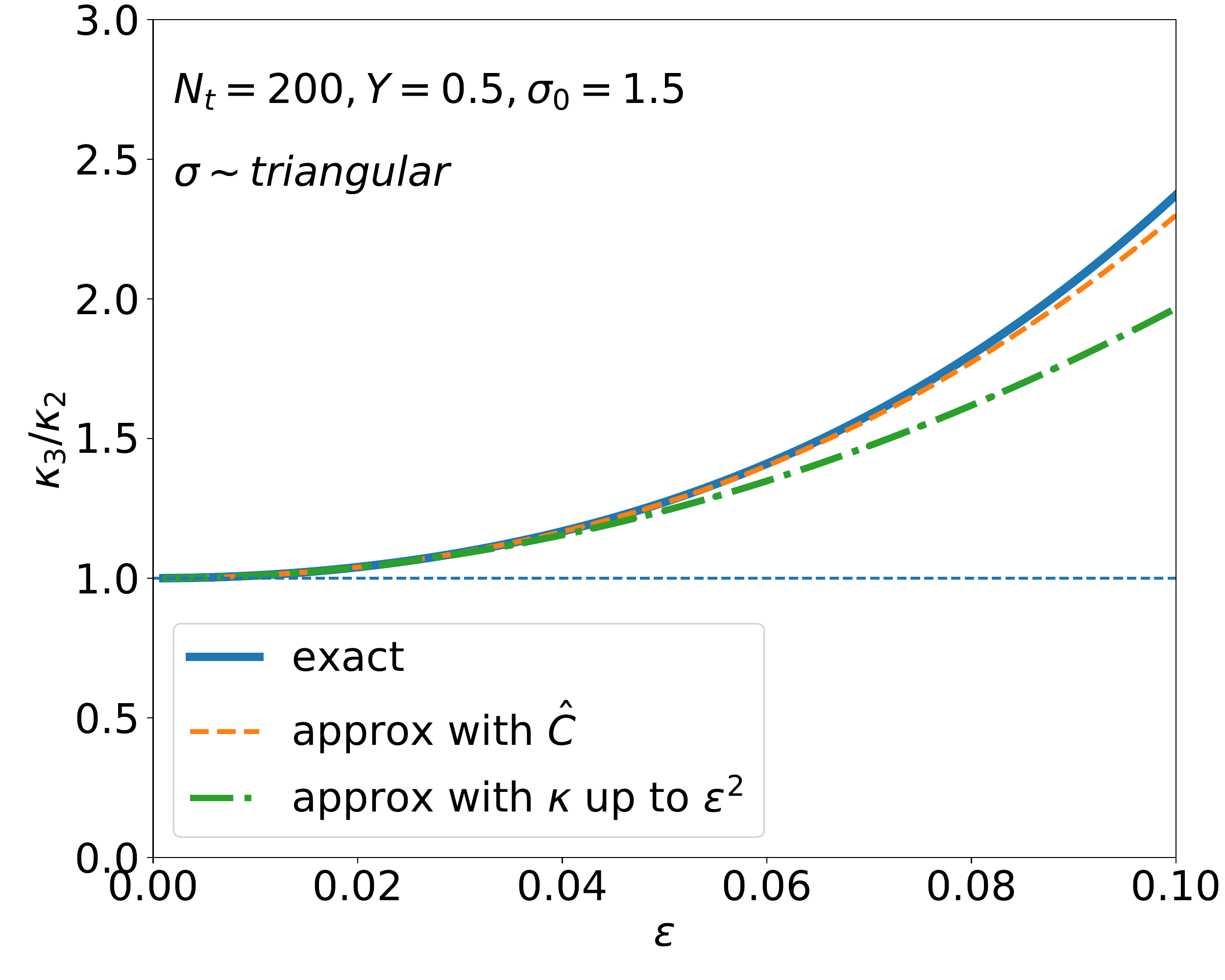}
    \includegraphics[width=0.46\textwidth]{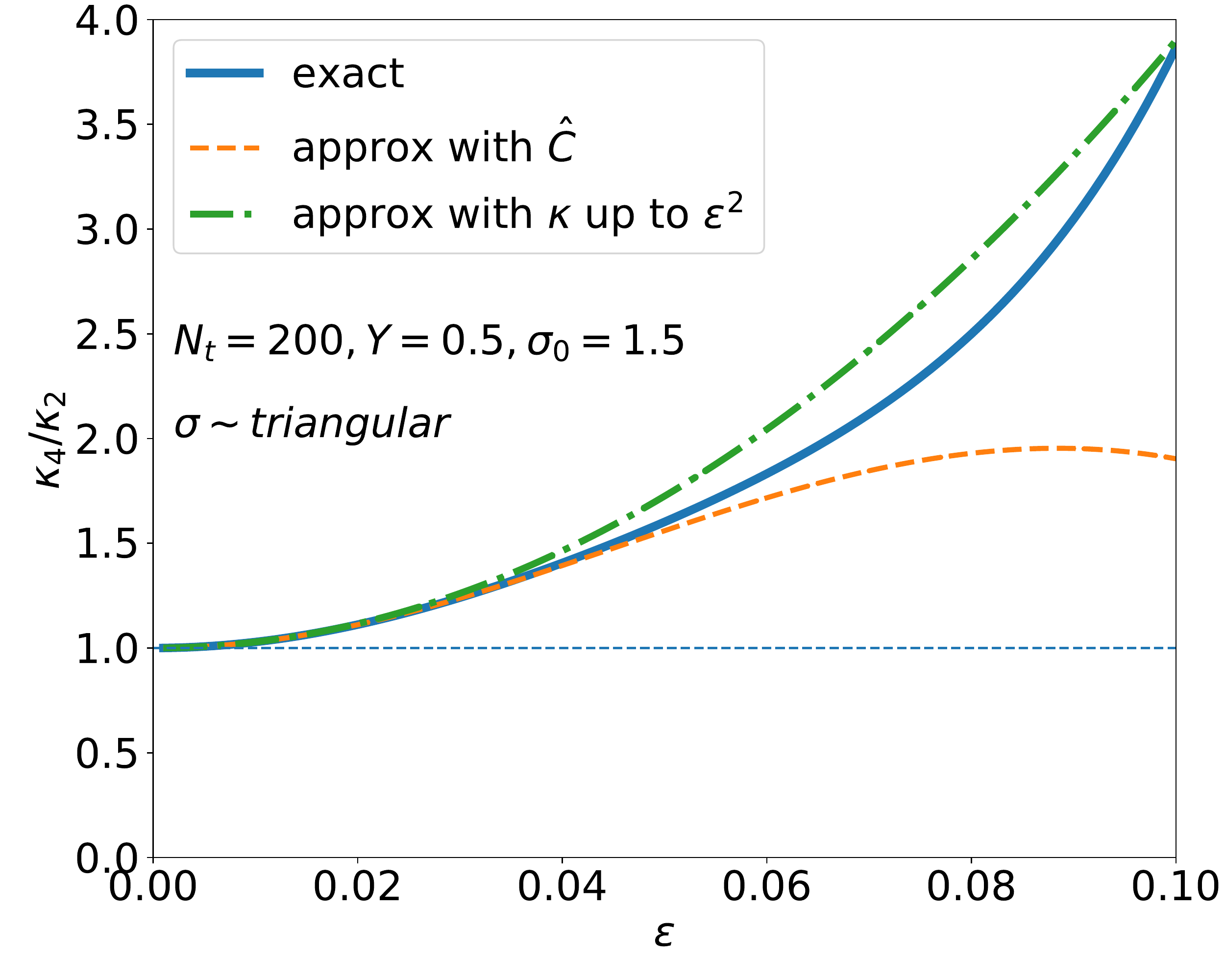}
    \includegraphics[width=0.46\textwidth]{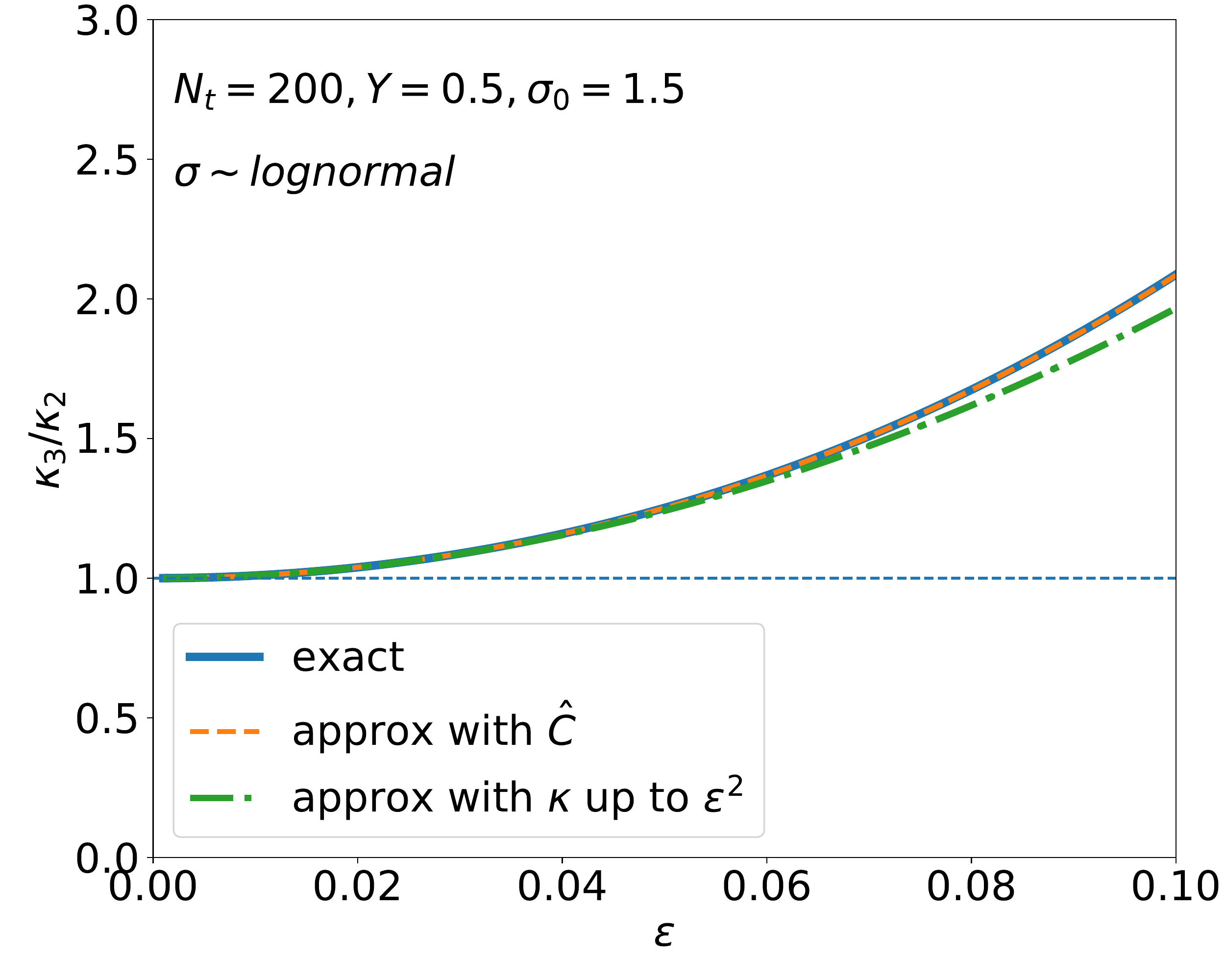}
    \includegraphics[width=0.46\textwidth]{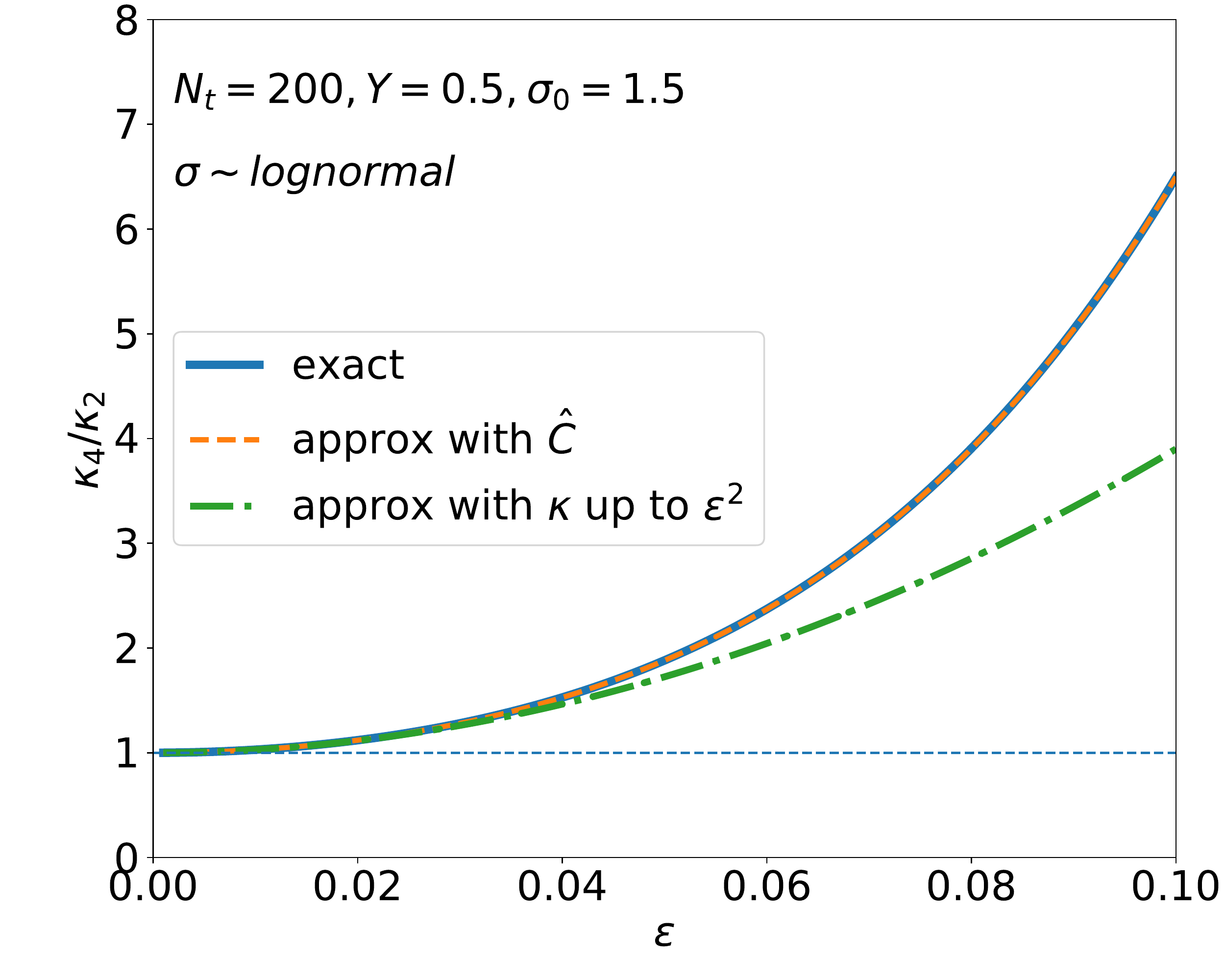}
\caption{The cumulant ratios, $\kappa_3/\kappa_2$ and $\kappa_4/\kappa_2$, calculated from the quadratic proton rapidity density distribution \eqref{eq:quadratic-rho} with $\msg$ fluctuations following uniform distribution \eqref{eq:unif} (first row), triangular distribution \eqref{eq:trian} (second row), and lognormal distribution \eqref{eq:lognorm} (third row). ``exact'' is the exact analytic result from Eqs. \eqref{eq:c1-quadr}, \eqref{eq:c2-quadr}, \eqref{eq:c3-quadr}, \eqref{eq:c4-quadr}, with $m_k$ given by Eq. \eqref{eq:mk-unif}, \eqref{eq:mk-trian}, and \eqref{eq:mk-logn}, respectively. ``approx with $\hat{C}$'' is the result using the leading-order terms in $\meps$ for the factorial cumulants (see main text). ``approx with $\kappa$ up to $\meps^2$'' is the approximation using Eq. \eqref{eq:cum-ratio-expansion}. The values of $N_t$, $Y$, and $\msg_0$ correspond to the STAR measurements in central Au+Au collisions at 7.7~GeV. \label{fig:cumulant-sig}}
\end{center}
\end{figure}

The cumulant ratios $\kappa_3/\kappa_2$ and $\kappa_4/\kappa_2$ with $\sigma$ following the uniform distribution are presented in the first row, the results with triangular distribution are in the second row, whereas the results with lognormal distribution are in the third row of Fig. \ref{fig:cumulant-sig}. The exact curves use the analytic exact results with all $A_i$ terms of the factorial cumulants. The dashed line $\kappa_n/\kappa_m = 1$ is the Poisson baseline for no correlations case. The cumulant ratios $\kappa_5/\kappa_2$ and $\kappa_6/\kappa_2$ are presented in Appendix \ref{app:c5-c6}.

\vfill
\newpage

In the case of the truncated normal distribution, we calculate the factorial cumulants with $N_t = 200$, $Y=0.5$, $\msg_0 = 1.5$ using numerical integration for $m_k$ (see previous section). The integration is done for $\msg \in [0.01, 3]$.\footnote{We have verified that both making this interval wider ($\msg \in [0.005, 5]$) or more narrow ($\msg \in [0.9, 2.1]$) does not change the cumulant ratios $\kappa_3/\kappa_2$ and $\kappa_4/\kappa_2$ for $\meps \in [0.01, 0.09]$. We have checked that the normalization constant is very close to 1 in all these intervals.}

\begin{figure}[H]
\centering
    \includegraphics[width=0.46\textwidth]{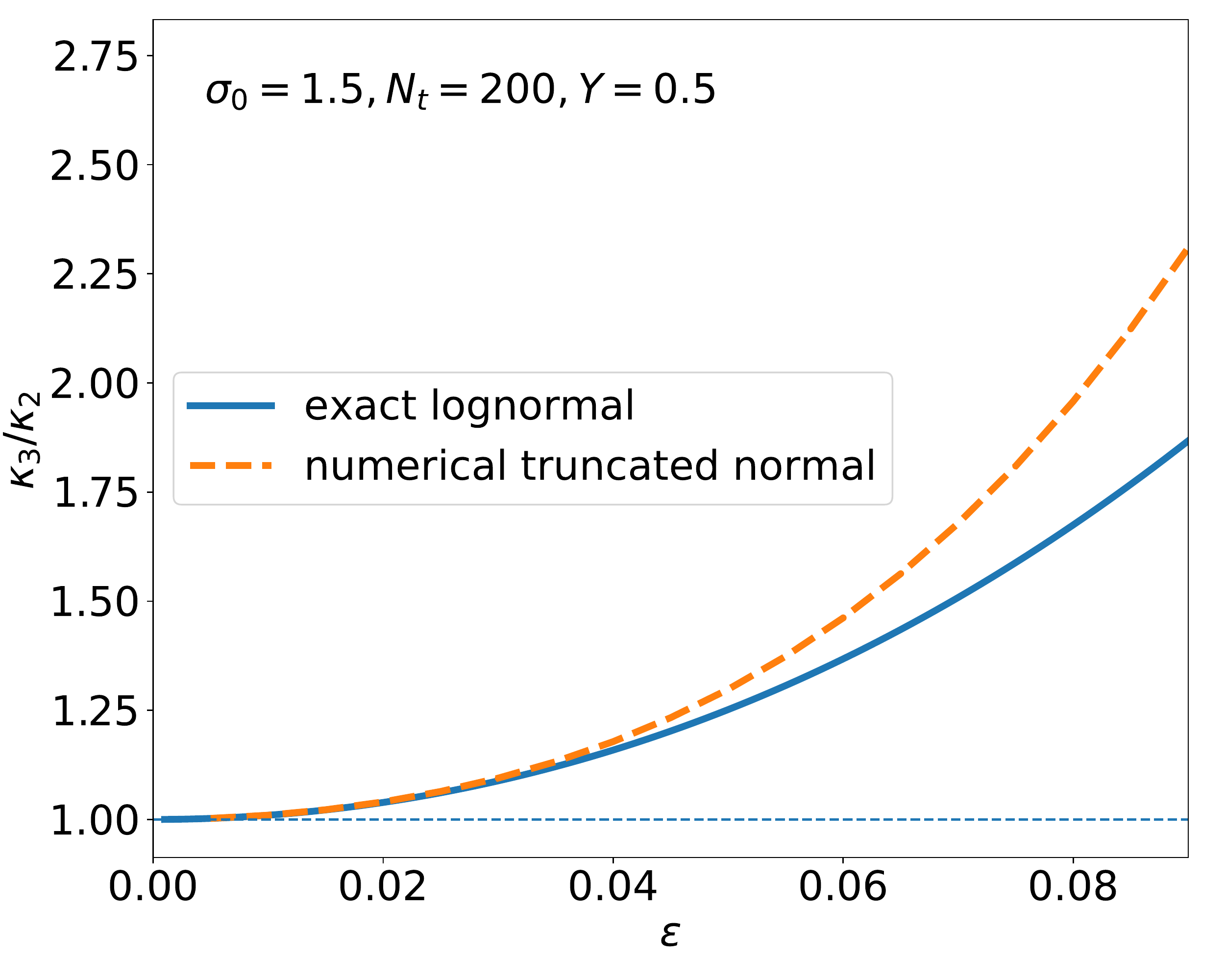}
    \includegraphics[width=0.46\textwidth]{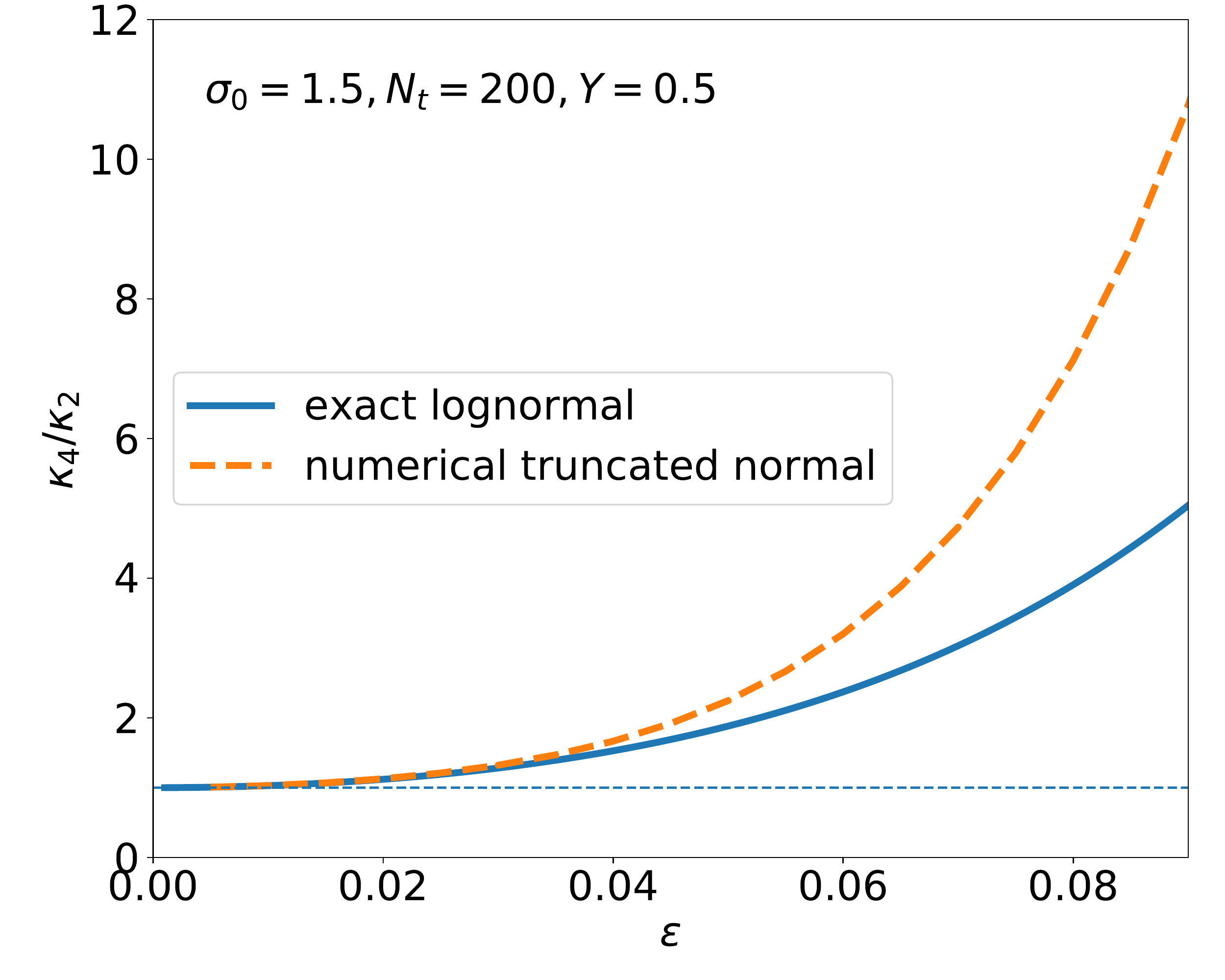}
\caption{The cumulant ratios, $\kappa_3/\kappa_2$ and $\kappa_4/\kappa_2$ calculated from the quadratic proton rapidity density distribution \eqref{eq:quadratic-rho}. The solid line assumes $\msg$ fluctuations following lognormal distribution (same as in Fig. \ref{fig:cumulant-sig}), whereas the dashed line corresponds to the truncated normal distribution with the $m_k$'s calculated numerically. The values of $N_t$, $Y$, and $\msg_0$ correspond to the STAR measurements in central Au+Au collisions at 7.7~GeV.}
\label{fig:cumulant-sig-norm}
\end{figure}

In Fig. \ref{fig:cumulant-sig-norm}, we show the numerical results for the truncated normal distribution in comparison to the exact results with the lognormal distribution. The normal distribution results are in the agreement with lognormal results for small $\meps$, and for greater $\meps$ they follow the same trend though they give greater values. 

\subsection{$\sqrt{s_{_{NN}}} = 3$~GeV}
We note that the cumulant ratios at $\sqrt{s_{_{NN}}}=3$~GeV are measured by the STAR Collaboration in the asymmetric rapidity range, $y \in [-0.5, 0]$, whereas, at other collision energies, they are obtained within the symmetric interval, $y \in [-0.5, 0.5]$ \cite{STAR:2021fge}. As seen, from Eqs. \eqref{eq:dn-dy-quadr-avg-sig}, \eqref{eq:corr2-quadr}, \eqref{eq:corr3-quadr}, and \eqref{eq:corr4-quadr}, the rapidity density distributions, as well as rapidity correlation functions are the even functions with respect to every $y_i$. Clearly, for any function $f$ satisfying this condition, 
\begin{equation} \label{eq:asym-vs-sym-integration}
\int_{-Y}^{Y} dy_1 \cdots \int_{-Y}^{Y} dy_n f(y_1, y_2, ..., y_n) = 2^n \int_{-Y}^{0} dy_1 \cdots \int_{-Y}^{0} dy_n f(y_1, y_2, ..., y_n)\,. 
\end{equation}
Therefore, the kth factorial cumulant from the rapidity correlation function in $[-Y, 0]$ is $2^k$ times smaller than the corresponding factorial cumulant in $[-Y, Y]$. Since the cumulants are the linear combinations of the factorial cumulants of a different order \cite{Friman:2022wuc}, the cumulant ratios are modified in a more complicated way.

Using $N_t = 175$ and $\msg_0 = 0.75$ extracted from the fit to the STAR data at $\sqrt{s_{_{NN}}} = 3$~GeV \cite{Kimelman:2023bmt} and taking Eq. \eqref{eq:asym-vs-sym-integration} into account, we have calculated the cumulant ratios, originating from the width fluctuations, in $y \in [-0.5, 0]$ as well as $y \in [-0.5, 0.5]$. They are presented in Fig. \ref{fig:cumulant-sig-yrange} for various $\msg$ distributions. The higher-order cumulant ratios are shown in Appendix \ref{app:c5-c6}. We note that the results obtained in symmetric and asymmetric rapidity intervals differ significantly. We have checked that the previously discussed two methods of approximation also work but in Eq. \eqref{eq:cum-ratio-expansion} $a_n$ should be divided by 2.
\begin{figure}[H]
\begin{center}
    \includegraphics[width=0.46\textwidth]{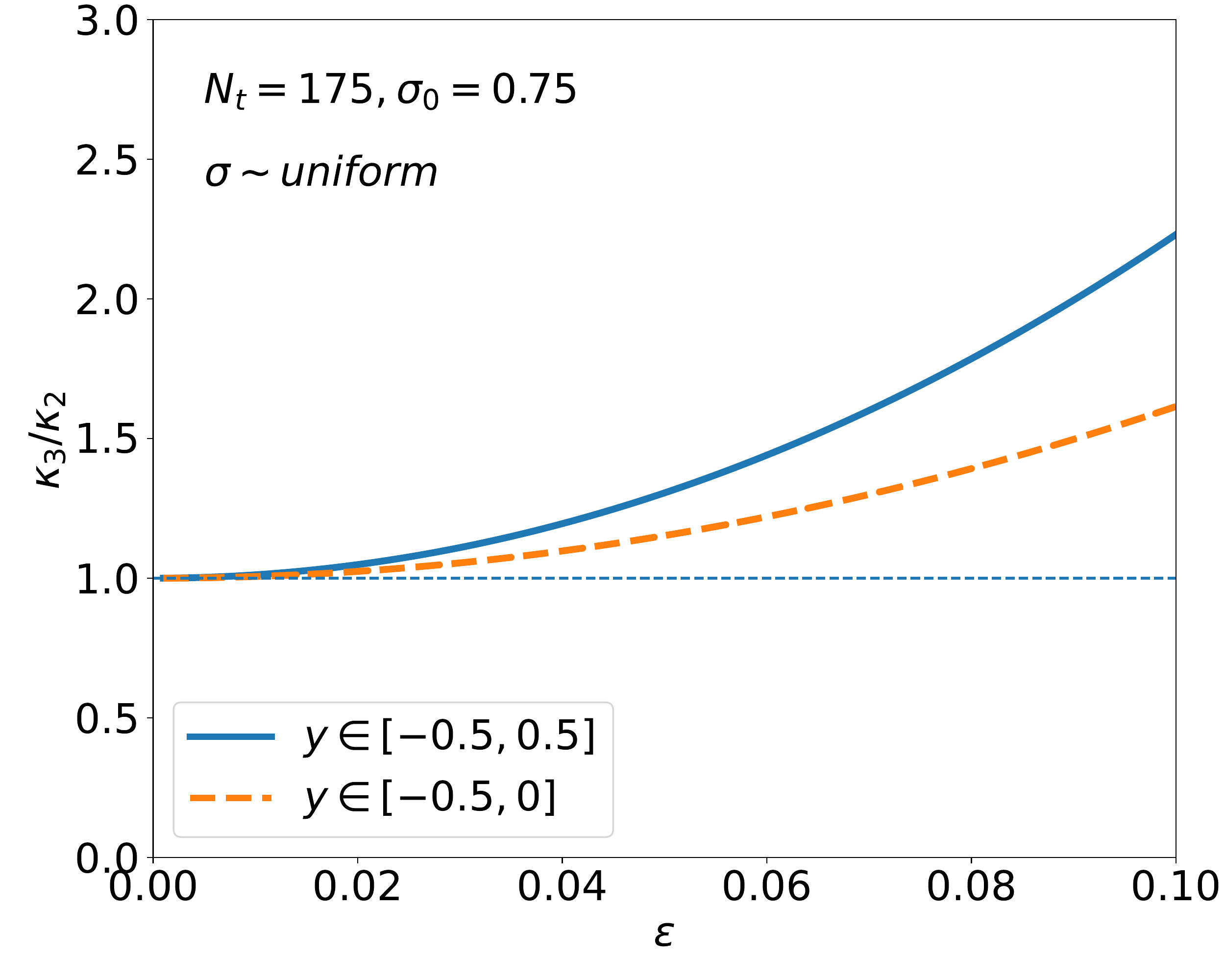}
    \includegraphics[width=0.46\textwidth]{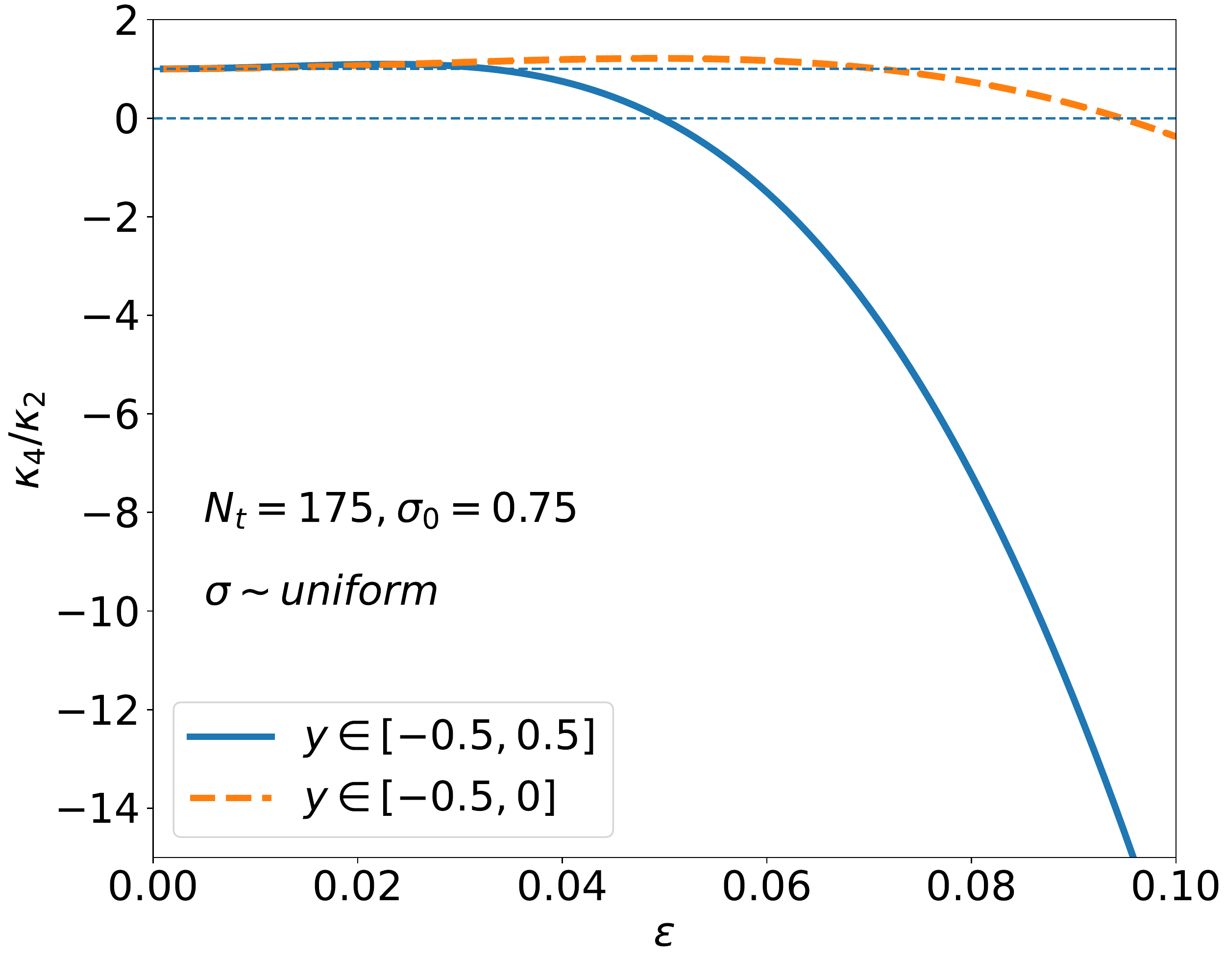}
    \includegraphics[width=0.46\textwidth]{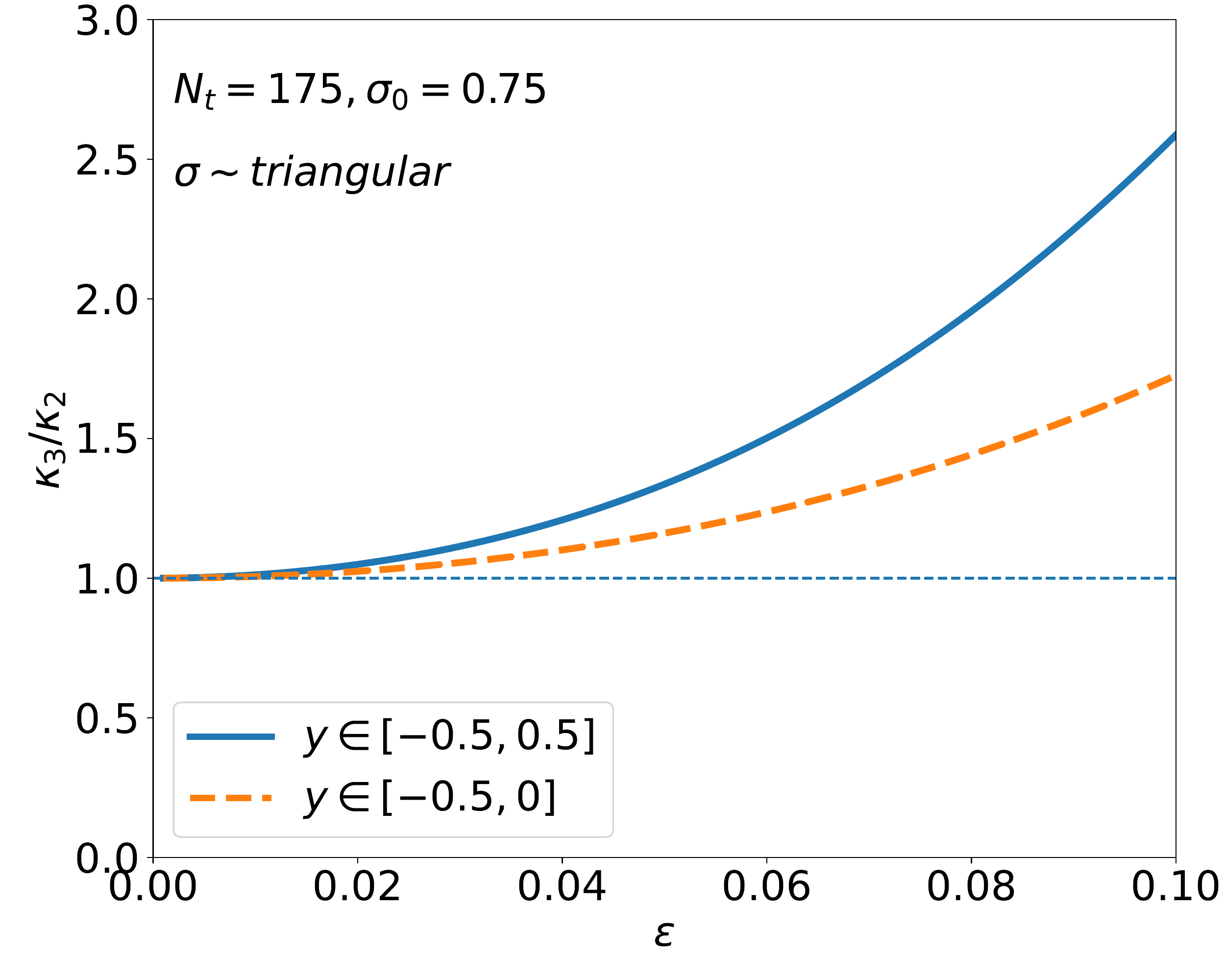}
    \includegraphics[width=0.46\textwidth]{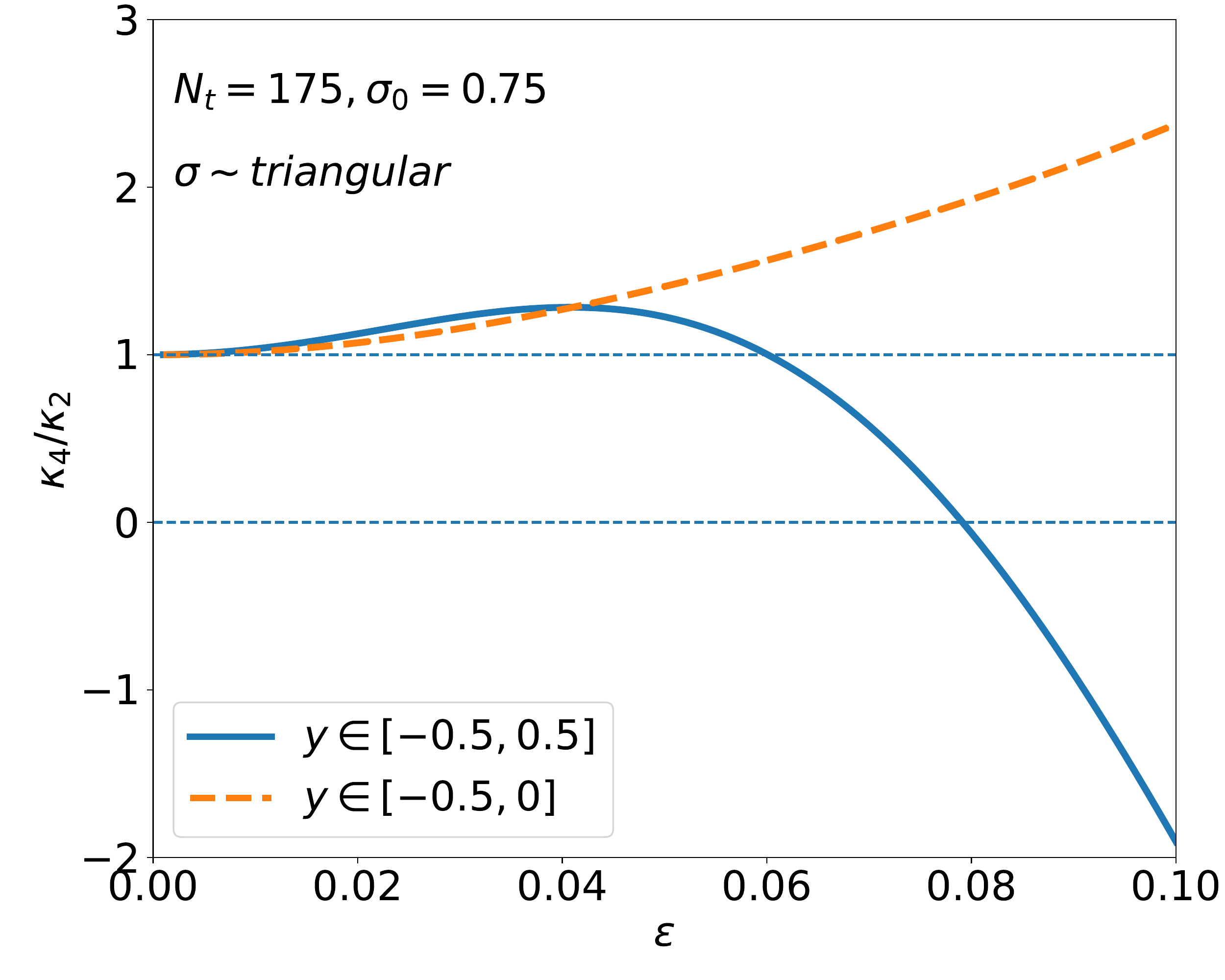}
    \includegraphics[width=0.46\textwidth]{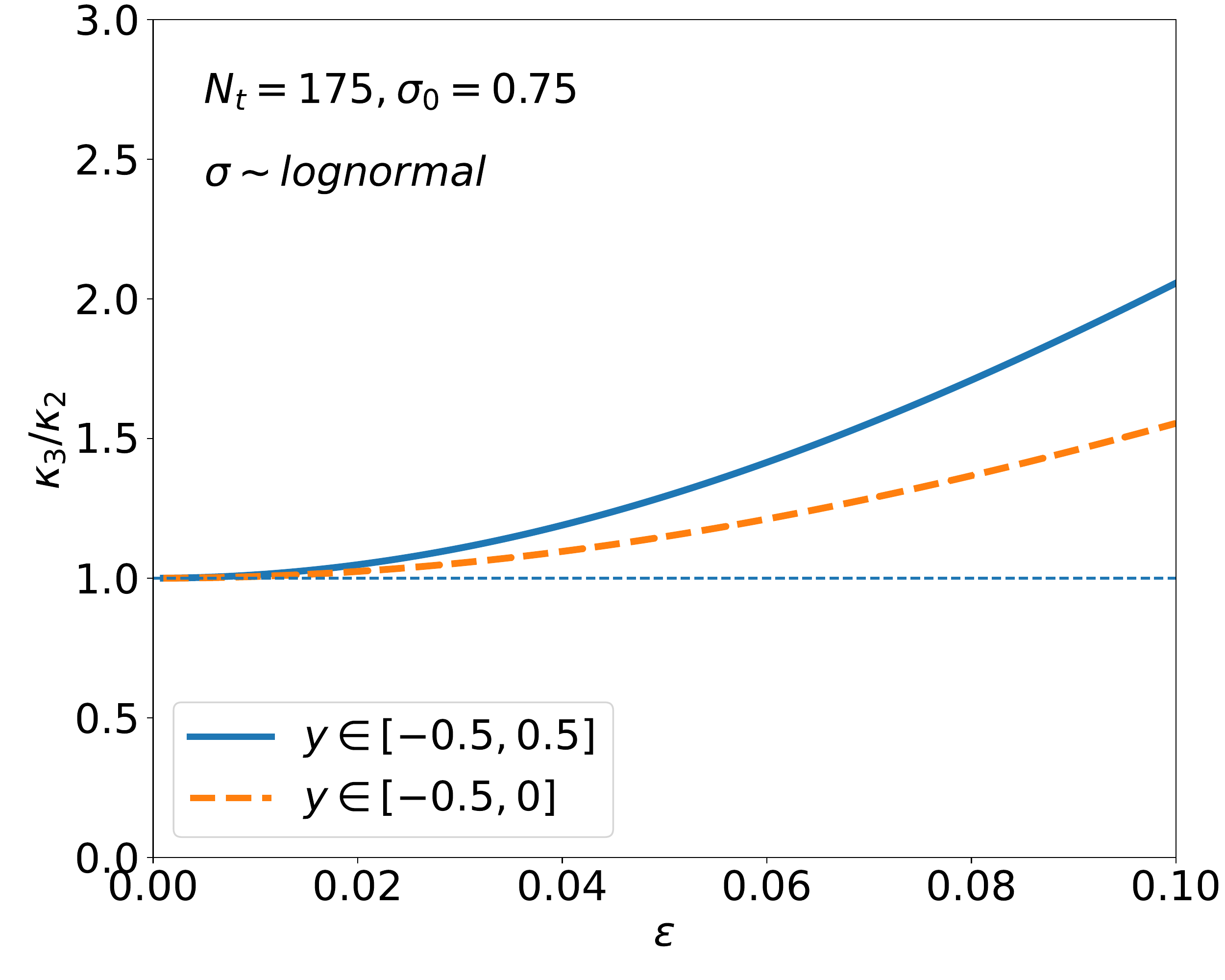}
    \includegraphics[width=0.46\textwidth]{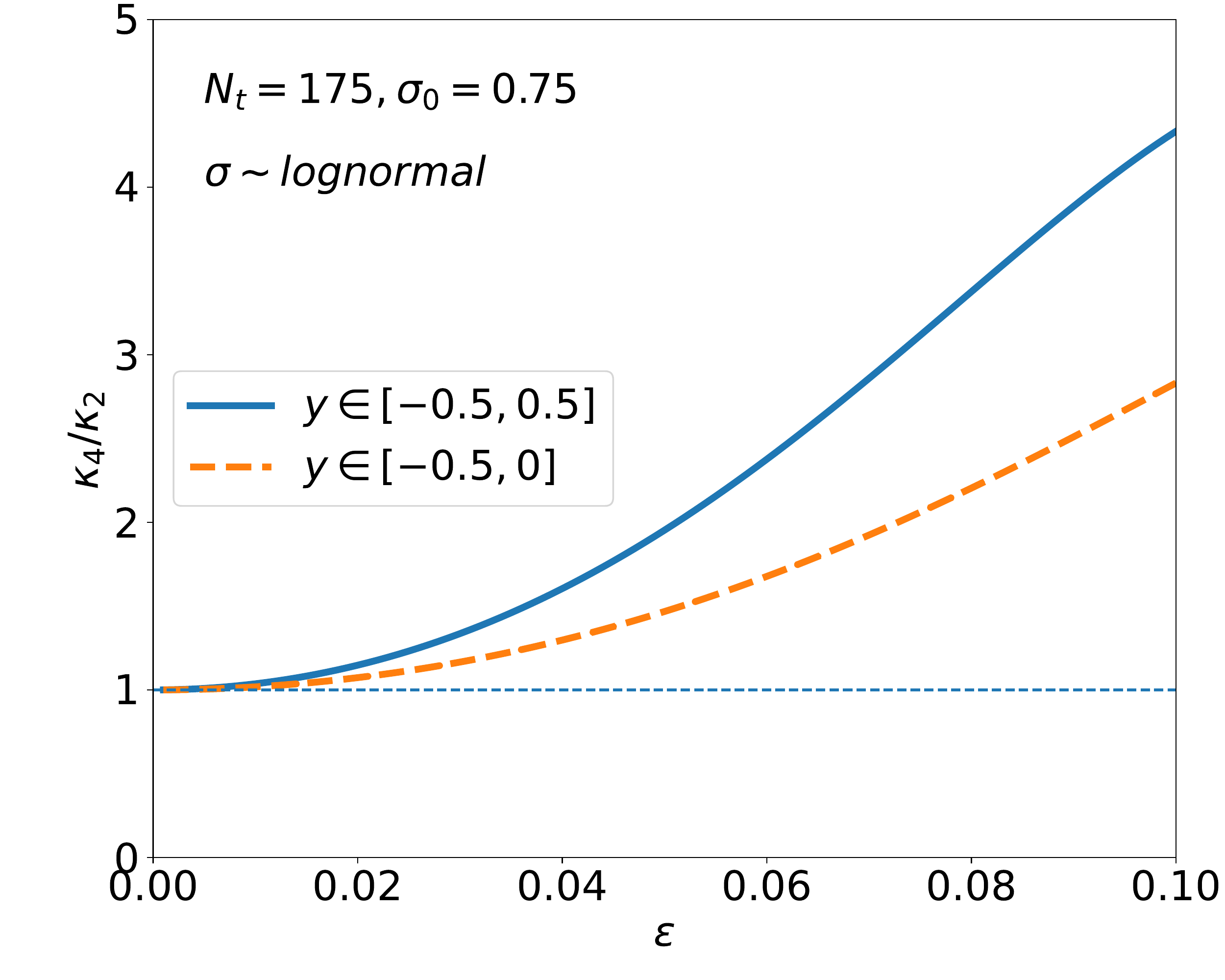}
\caption{The cumulant ratios, $\kappa_3/\kappa_2$ and $\kappa_4/\kappa_2$ calculated from the quadratic proton rapidity density distribution \eqref{eq:quadratic-rho} with $\msg$ following uniform (first row), triangular (second row), and lognormal (third row) distribution using exact formulas and values $N_t$, $\msg_0$ corresponding to central Au+Au collisions at 3~GeV. The results are compared for the symmetric ($y \in [-0.5, 0.5]$) and asymmetric ($y \in [-0.5, 0]$) rapidity range.\label{fig:cumulant-sig-yrange}}
\end{center}
\end{figure}

\section{Discussion and summary}
In this paper, we have presented the method of analytical calculation of the correlation functions, factorial cumulants, and cumulants originating from the fluctuating width of the proton rapidity density distribution. We have assumed that this distribution is a Gaussian function that can be approximated by the quadratic function in the midrapidity region. This assumption is valid for central low-energy Au+Au collisions. At higher energies (already about 10~GeV), the distribution has a bimodal shape. We have studied the width, $\msg$, fluctuations following three qualitatively different distributions, uniform, triangular, and lognormal (which approximates normal distribution). Obviously, they do not expend all the possibilities but we believe they constitute a representative choice. 

We find that for very different $\msg$ distributions, all the cumulant ratios can be approximated by the universal function \eqref{eq:cum-ratio-expansion} for small $\meps$, where $\meps$ determines the strength of the $\msg$ fluctuations, see Eq. (\ref{eq:std-constr}). This indicates that the results are independent of a choice of $p(\msg)$ when the width fluctuations are small. This approximation is equivalent to considering only the leading-order terms of the first and second factorial cumulants. Hence, at small $\meps$, the cumulant ratios are governed by the two-particle correlations.

In Tab. \ref{tab:star}, we present the STAR Collaboration data. We note that the cumulant ratios from the width fluctuations for $\meps < 0.05$ are of the same order of magnitude as measured experimentally. Hence, we have shown that the distribution width fluctuations due to fireball density fluctuations may have a measurable contribution to the cumulant ratios in heavy-ion experiments. Therefore, they should be further studied and taken into account in the search for the predicted first-order phase transition and critical point between the hadronic matter and quark-gluon plasma. A direct comparison with data is rather challenging because there are other sources of correlations, e.g., the global baryon number conservation \cite{ALICE:2019nbs,Braun-Munzinger:2016yjz,Skokov:2012ds,Bzdak:2012an,Braun-Munzinger:2020jbk,Vovchenko:2020tsr,Barej:2020ymr,Barej:2022jij}. It would be interesting to investigate this effect using various Monte Carlo models.
\begin{table}[H]
\begin{center}
\caption{Cumulant ratios measured by the STAR Collaboration at the lowest energies \cite{STAR:2023ndx}. \label{tab:star}}
\begin{tabular}{|r|r|r|}
\hline
$\sqrt{s_{NN}}$ & 3~GeV & 7.7~GeV \\ \hline
 & $y \in [-0.5, 0]$ & $y \in [-0.5, 0.5]$ \\ \hline
$\kappa_3/\kappa_2$ & $\approx0.95 \pm 0.05$ & $\approx0.83 \pm 0.07$ \\ \hline
$\kappa_4/\kappa_2$ & $\approx -0.8 \pm 0.8$ & $\approx1.8 \pm 1.1$ \\ \hline
\end{tabular}
\end{center}
\end{table}

We have also demonstrated that the cumulant ratios measured in a symmetric and an asymmetric rapidity range might result in very different values. Therefore, one should be very careful when comparing the cumulant ratios measured in different rapidity intervals.

We note that the longitudinal fluctuations have already been studied by the expansion into orthogonal polynomials \cite{Bzdak:2012tp,Bzdak:2015dja,Jia:2015jga,ATLAS:2016rbh}. In this formalism, the width fluctuation is reflected in the $a_2$ coefficient. In this paper, we have proposed another approach that focuses directly on the rapidity density distribution width fluctuations originating from the longitudinal fluctuations of the fireball density. 

The goal of this work is to stimulate further research of longitudinal fluctuations. In principle, it can not only contribute to the important probes of a critical point such as the proton multiplicity cumulants but also improve our understanding of the fireball longitudinal dynamics.

\begin{acknowledgements}
This work was partially supported by the Ministry of Science and Higher Education (PL), and by the National Science Centre (PL), Grant No. 2018/30/Q/ST2/00101.
\end{acknowledgements}


\appendix 
\section{The higher-order correlation functions and factorial cumulants} \label{app:c5-c6}
The subsequent $n$-particle rapidity density distributions follow the same pattern ($n=5,6$) as in Eqs. \eqref{eq:rho2}, \eqref{eq:rho3}, and \eqref{eq:rho4}. Namely,
\begin{equation} \label{eq:rho5}
\begin{split}
\mrho_{\text{meas},5}(y_1, y_2, ..., y_5) = \ml(\mf{N_t}{\sqrt{2\pi}}\mr)^5 &\ml[ m_5 -\mf{1}{2}m_7 \sum_{i=1}^{5} y_i^2 + \mf{1}{4} m_9 \sum_{j>i} y_i^2 y_j^2 - \mf{1}{8} m_{11} \sum_{k>j>i} y_i^2 y_j^2 y_k^2 \mr. \\
&\left.+ \mf{1}{16} m_{13} \sum_{l>k>j>i} y_i^2 y_j^2 y_k^2 y_l^2 - \mf{1}{32}m_{15} y_1^2 y_2^2 \cdots y_5^2 \mr] \,,
\end{split}
\end{equation}
\begin{equation} \label{eq:rho6}
\begin{split}
\mrho_{\text{meas},6}(y_1, y_2, ..., y_6) = \ml(\mf{N_t}{\sqrt{2\pi}}\mr)^6 &\ml[ m_6 -\mf{1}{2}m_8 \sum_{i=1}^{6} y_i^2 + \mf{1}{4} m_{10} \sum_{j>i} y_i^2 y_j^2 - \mf{1}{8} m_{12} \sum_{k>j>i} y_i^2 y_j^2 y_k^2 \mr. \\
&+ \mf{1}{16} m_{14} \sum_{l>k>j>i} y_i^2 y_j^2 y_k^2 y_l^2 - \mf{1}{32}m_{16} \sum_{n>l>k>j>i} y_i^2 y_j^2 y_k^2 y_l^2 y_n^2\\ 
&\left.+ \mf{1}{64}m_{18} y_1^2 y_2^2 \cdots y_6^2 \mr] \,,
\end{split}
\end{equation}

The five-particle correlation function and the corresponding factorial cumulant read\footnote{They are calculated according to formulas given in Ref. \cite{Bzdak:2015dja}.}
\begin{equation}
\begin{split}
C_5(y_1, y_2, ..., y_5) = \ml(\mf{N_t}{\sqrt{2\pi}}\mr)^5 \mf{1}{2^5} &\ml[32 A_0 -16 A_1 \sum_{i=1}^{5} y_i^2 +8 A_2 \sum_{j>i} y_i^2 y_j^2 - 4 A_3 \sum_{k>j>i} y_i^2 y_j^2 y_k^2 \mr. \\
 &\ml. -2 A_4 \sum_{l>k>j>i} y_i^2 y_j^2 y_k^2 y_l^2 - A_5 \: y_1^2 y_2^2 y_3^2 y_4^2 y_5^2  \mr] \,,
\end{split}
\end{equation}
\begin{equation}
\hat{C}_5 = \mf{N_t^5 Y^5}{(2 \pi)^{5/2}} \ml[32 A_0 - \mf{80}{3} A_1 Y^2 + \mf{80}{9} A_2 Y^4 - \mf{40}{27} A_3 Y^6 - \mf{10}{81} A_4 Y^8 - \mf{1}{243} A_5 Y^{10} \mr] \,,
\end{equation}
where
\begin{equation}
\begin{split}
A_0 &= 24 m_1^5 - 60 m_1^3 m_2 + 30 m_1 m_2^2 + 20 m_1^2 m_3 - 10 m_2 m_3 - 5 m_1 m_4 + m_5 \,, \\
A_1 &= 24 m_1^4 m_3 - 36 m_1^2 m_2 m_3 + 6 m_2^2 m_3 + 8 m_1 m_3^2 - 24 m_1^3 m_4 + 24 m_1 m_2 m_4 - 5 m_3 m_4 + 12 m_1^2 m_5 \\
&\phantom{=} - 6 m_2 m_5 - 4 m_1 m_6 + m_7 \,, \\
A_2 &= 24 m_1^3 m_3^2 - 18 m_1 m_2 m_3^2 + 2 m_3^3 - 36 m_1^2 m_3 m_4 + 12 m_2 m_3 m_4 + 12 m_1 m_4^2 + 12 m_1 m_3 m_5 \\
&\phantom{=} - 6 m_4 m_5  - 6 m_1^3 m_6 + 6 m_1 m_2 m_6 - 3 m_3 m_6 + 6 m_1^2 m_7 - 3 m_2 m_7 - 3 m_1 m_8 + m_9 \,, \\
A_3 &= -2 m_1 m_{10} + m_{11} + 24 m_1^2 m_3^3 - 6 m_2 m_3^3 - 36 m_1 m_3^2 m_4 +  12 m_3 m_4^2 + 6 m_3^2 m_5 - 18 m_1^2 m_3 m_6  \\
&\phantom{=} + 6 m_2 m_3 m_6 + 12 m_1 m_4 m_6 -  3 m_5 m_6 + 12 m_1 m_3 m_7 - 6 m_4 m_7 - 3 m_3 m_8 + 2 m_1^2 m_9 - m_2 m_9\,, \\
A_4 &= m_1 m_{12} - m_{13} + 4 m_{10} m_3 - 24 m_1 m_3^4 + 24 m_3^3 m_4 + 36 m_1 m_3^2 m_6 - 24 m_3 m_4 m_6 - 6 m_1 m_6^2 \\
&\phantom{=} - 12 m_3^2 m_7 + 6 m_6 m_7 - 8 m_1 m_3 m_9 + 4 m_4 m_9\,, \\
A_5 &= m_{15} - 5 m_{12} m_{3} + 24 m_{3}^5 - 60 m_{3}^3 m_6 + 30 m_{3} m_6^2 + 20 m_{3}^2 m_9 - 10 m_6 m_9\,.
\end{split}
\end{equation}

The six-particle correlation function and the factorial cumulant are given by
\begin{equation}
\begin{split}
C_6(y_1, y_2, ..., y_6) = &\ml(\mf{N_t}{\sqrt{2\pi}}\mr)^6 \mf{1}{2^6} \ml[-64 A_0 +32 A_1 \sum_{i=1}^{6} y_i^2 +16 A_2 \sum_{j>i} y_i^2 y_j^2 +8 A_3 \sum_{k>j>i} y_i^2 y_j^2 y_k^2  \mr. \\
 &\ml.+4 A_4 \sum_{l>k>j>i} y_i^2 y_j^2 y_k^2 y_l^2 +2 A_5 \sum_{n>l>k>j>i} y_i^2 y_j^2 y_k^2 y_l^2 y_n^2 + A_6\: y_1^2 y_2^2 \cdots y_6^2    \mr] \,,
\end{split}
\end{equation}
\begin{equation}
\hat{C}_6 = \mf{N_t^6 Y^6}{(2 \pi)^{3}} \ml[-64 A_0 +64 A_1 Y^2 + \mf{80}{3} A_2 Y^4 + \mf{160}{27} A_3 Y^6 + \mf{20}{27} A_4 Y^8 + \mf{4}{81} A_5 Y^{10} + \mf{1}{729} A_6 Y^{12} \mr] \,,
\end{equation}
\vfill
\newpage
where
\begin{equation}
\begin{split}
A_0 &= 120 m_1^6 - 360 m_1^4 m_2 + 270 m_1^2 m_2^2 - 30 m_2^3 + 120 m_1^3 m_3 - 120 m_1 m_2 m_3 + 10 m_3^2 - 30 m_1^2 m_4  \\
&\phantom{=} + 15 m_2 m_4 + 6 m_1 m_5 - m_6 \,, \\
A_1 &= 120 m_1^5 m_3 - 240 m_1^3 m_2 m_3 + 90 m_1 m_2^2 m_3 + 60 m_1^2 m_3^2 -  20 m_2 m_3^2 - 120 m_1^4 m_4 + 180 m_1^2 m_2 m_4 \\
&\phantom{=} - 30 m_2^2 m_4 -  50 m_1 m_3 m_4 + 5 m_4^2 + 60 m_1^3 m_5 - 60 m_1 m_2 m_5 + 11 m_3 m_5 -  20 m_1^2 m_6 + 10 m_2 m_6 \\
&\phantom{=} + 5 m_1 m_7 - m_8 \,, \\
A_2 &= m_{10} - 120 m_1^4 m_3^2 + 144 m_1^2 m_2 m_3^2 - 18 m_2^2 m_3^2 - 24 m_1 m_3^3 +  192 m_1^3 m_3 m_4 - 144 m_1 m_2 m_3 m_4 \\
&\phantom{=} + 18 m_3^2 m_4 - 72 m_1^2 m_4^2 +  24 m_2 m_4^2 - 72 m_1^2 m_3 m_5 + 24 m_2 m_3 m_5 + 48 m_1 m_4 m_5 - 6 m_5^2 \\
&\phantom{=} +  24 m_1^4 m_6 - 36 m_1^2 m_2 m_6 + 6 m_2^2 m_6 + 24 m_1 m_3 m_6 - 9 m_4 m_6 -  24 m_1^3 m_7 + 24 m_1 m_2 m_7 \\
&\phantom{=}- 6 m_3 m_7 + 12 m_1^2 m_8 - 6 m_2 m_8 - 4 m_1 m_9 \,, \\
A_3 &= -6 m_1^2 m_{10} + 3 m_1 m_{11} - m_{12} + 3 m_{10} m_2 + 120 m_1^3 m_3^3 -  72 m_1 m_2 m_3^3 + 6 m_3^4 - 216 m_1^2 m_3^2 m_4 \\
&\phantom{=}+ 54 m_2 m_3^2 m_4 +  108 m_1 m_3 m_4^2 - 12 m_4^3 + 54 m_1 m_3^2 m_5 - 36 m_3 m_4 m_5 -  72 m_1^3 m_3 m_6  \\
&\phantom{=}+54 m_1 m_2 m_3 m_6 - 12 m_3^2 m_6 + 54 m_1^2 m_4 m_6 -  18 m_2 m_4 m_6 - 18 m_1 m_5 m_6 + 3 m_6^2 + 54 m_1^2 m_3 m_7 \\
&\phantom{=}- 18 m_2 m_3 m_7 -  36 m_1 m_4 m_7 + 9 m_5 m_7 - 18 m_1 m_3 m_8 + 9 m_4 m_8 + 6 m_1^3 m_9 -  6 m_1 m_2 m_9 \\
&\phantom{=}+ 4 m_3 m_9\,, \\
A_4 &= 2 m_1^2 m_{12} - 2 m_1 m_{13} + m_{14} - m_{12} m_2 + 16 m_1 m_{10} m_3 - 4 m_{11} m_3 -  120 m_1^2 m_3^4 + 24 m_2 m_3^4\\
&\phantom{=} - 8 m_{10} m_4 + 192 m_1 m_3^3 m_4 -  72 m_3^2 m_4^2 - 24 m_3^3 m_5 + 144 m_1^2 m_3^2 m_6 - 36 m_2 m_3^2 m_6\\
&\phantom{=} -  144 m_1 m_3 m_4 m_6 + 24 m_4^2 m_6 + 24 m_3 m_5 m_6 - 18 m_1^2 m_6^2 +  6 m_2 m_6^2 - 72 m_1 m_3^2 m_7\\
&\phantom{=} + 48 m_3 m_4 m_7 + 24 m_1 m_6 m_7 - 6 m_7^2 +  12 m_3^2 m_8 - 6 m_6 m_8 - 24 m_1^2 m_3 m_9 + 8 m_2 m_3 m_9\\
&\phantom{=} + 16 m_1 m_4 m_9 -  4 m_5 m_9\,, \\
A_5 &= m_1 m_{15} - m_{16} - 10 m_1 m_{12} m_3 + 5 m_{13} m_3 - 20 m_{10} m_3^2 + 120 m_1 m_3^5 +  5 m_{12} m_4 - 120 m_3^4 m_4 \\
&\phantom{=} + 10 m_{10} m_6 - 240 m_1 m_3^3 m_6 +  180 m_3^2 m_4 m_6 + 90 m_1 m_3 m_6^2 - 30 m_4 m_6^2 + 60 m_3^3 m_7 \\
&\phantom{=} -  60 m_3 m_6 m_7 + 60 m_1 m_3^2 m_9 - 40 m_3 m_4 m_9 - 20 m_1 m_6 m_9 + 10 m_7 m_9\,, \\
A_6 &= m_{18} - 6 m_{15} m_3 + 30 m_{12} m_3^2 - 120 m_3^6 - 15 m_{12} m_6 + 360 m_3^4 m_6 - 270 m_3^2 m_6^2 + 30 m_6^3 \\
&\phantom{=}- 120 m_3^3 m_9 + 120 m_3 m_6 m_9 - 10 m_9^2\,.
\end{split}
\end{equation}

In the case of the uniform $\msg$ distribution, Eq. \eqref{eq:unif}, the approximated formulas (leading order terms of the power series expansion about $\meps=0$) read:
\begin{equation}
\hat{C}_5 \approx - \ml(\mf{N_t z}{\sqrt{2 \pi}}\mr)^5 \mf{192 \meps^6}{7} \ml(2 - z^2  \mr)^4 \ml(1 - z^2 \mr)\,,
\end{equation}
\begin{equation}
\hat{C}_6 \approx \ml(\mf{N_t z}{\sqrt{2 \pi}}\mr)^6 \mf{48\meps^6}{7} \ml(2 - z^2  \mr)^6 \,.
\end{equation}

For the triangular distribution, Eq. \eqref{eq:trian}, we have:
\begin{equation}
\hat{C}_5 \approx - \ml(\mf{N_t z}{\sqrt{2 \pi}}\mr)^5 \mf{216\meps^6}{7} \ml(2 - z^2  \mr)^4 \ml(1 - z^2 \mr)\,,
\end{equation}
\begin{equation}
\hat{C}_6 \approx \ml(\mf{N_t z}{\sqrt{2 \pi}}\mr)^6 \mf{12\meps^6}{7} \ml(2 - z^2  \mr)^6 \,.
\end{equation}

For the lognormal distribution, Eq. \eqref{eq:lognorm}, we obtain:
\begin{equation}
\hat{C}_5 \approx \ml( \mf{N_t z}{\sqrt{2 \pi}} \mr)^5 5\meps^8 \ml(2 - z^2 \mr)^2 \ml(200 - 1172 z^2 + 1694 z^4 -675 z^6 \mr)\,,
\end{equation}
\begin{equation}
\hat{C}_6 \approx \ml( \mf{N_t z}{\sqrt{2 \pi}} \mr)^6 48\meps^{10} \ml(2 - z^2 \mr)^2 \ml(432 - 3664 z^2 + 8728 z^4 -7636 z^6 + 2187 z^8 \mr)\,.
\end{equation}

\begin{figure}[H]
\begin{center}
    \includegraphics[width=0.46\textwidth]{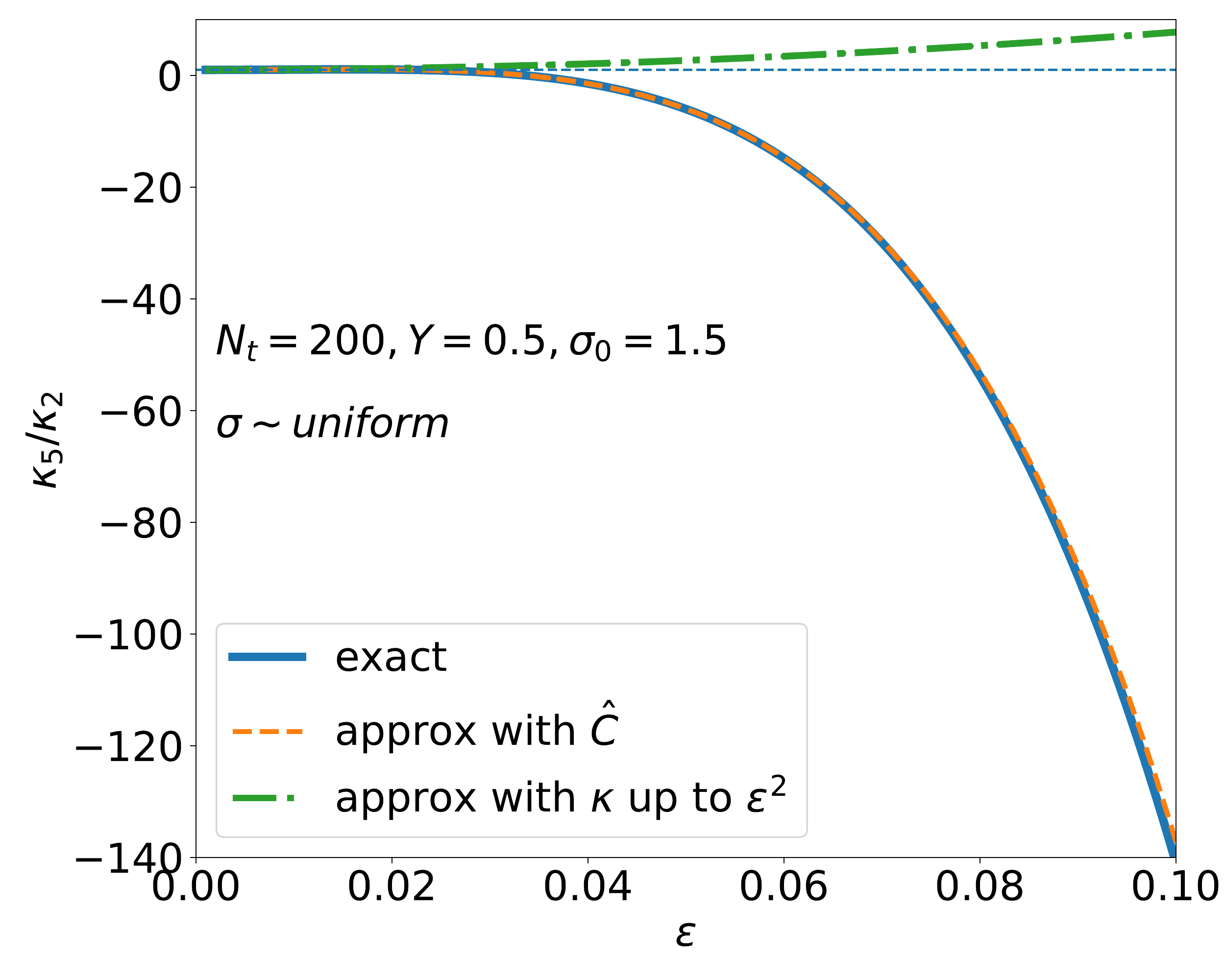}
    \includegraphics[width=0.46\textwidth]{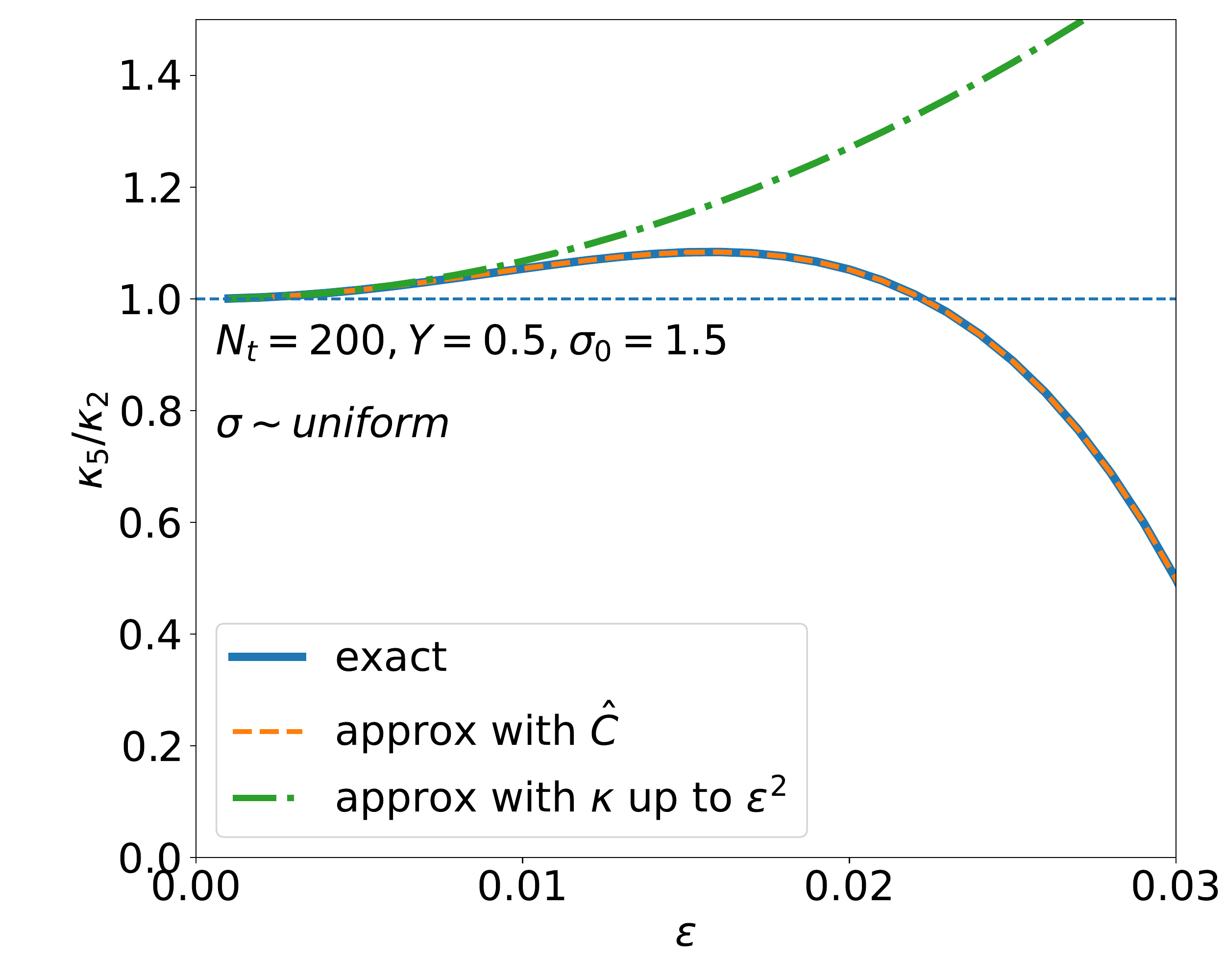}
    \includegraphics[width=0.46\textwidth]{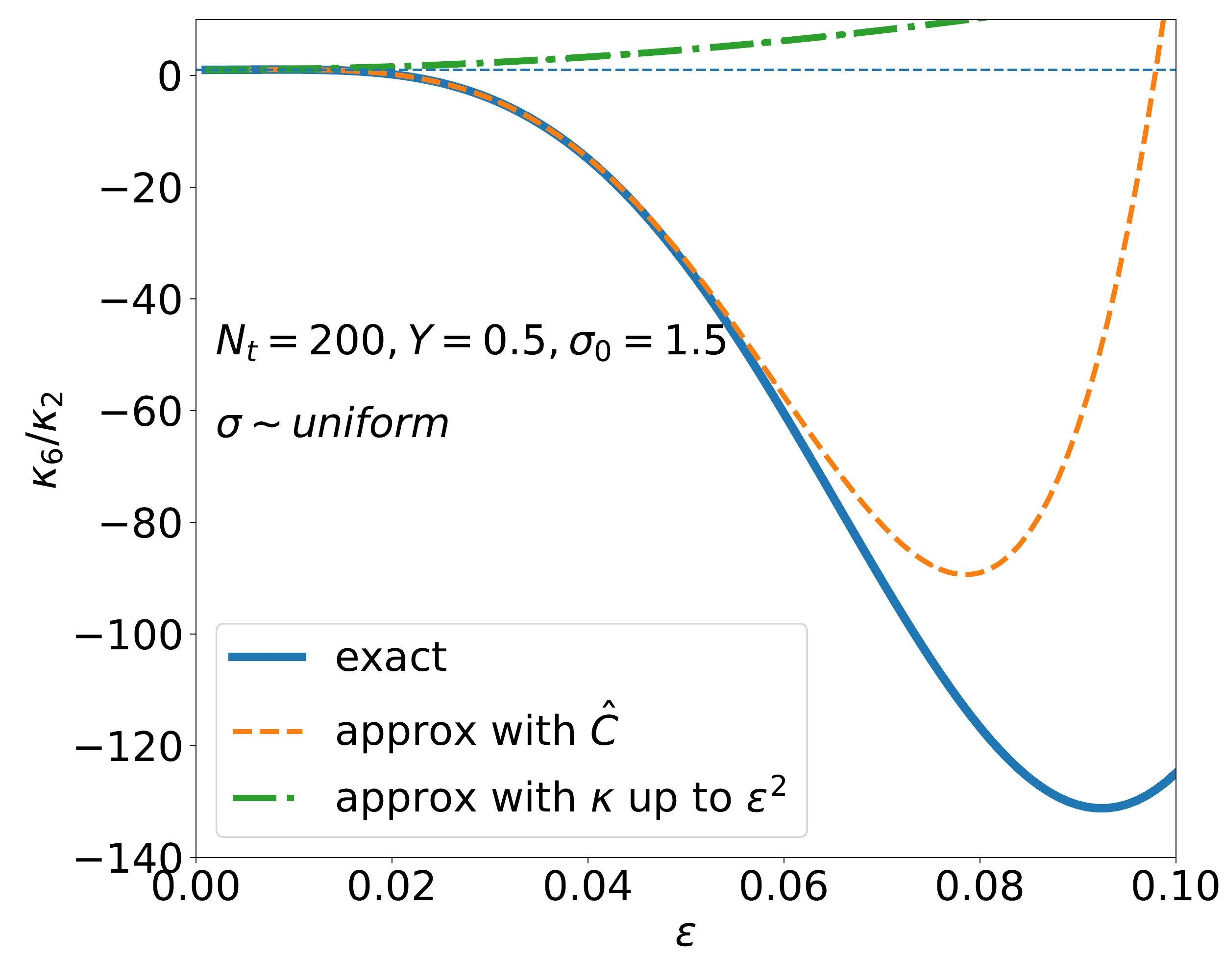}
    \includegraphics[width=0.46\textwidth]{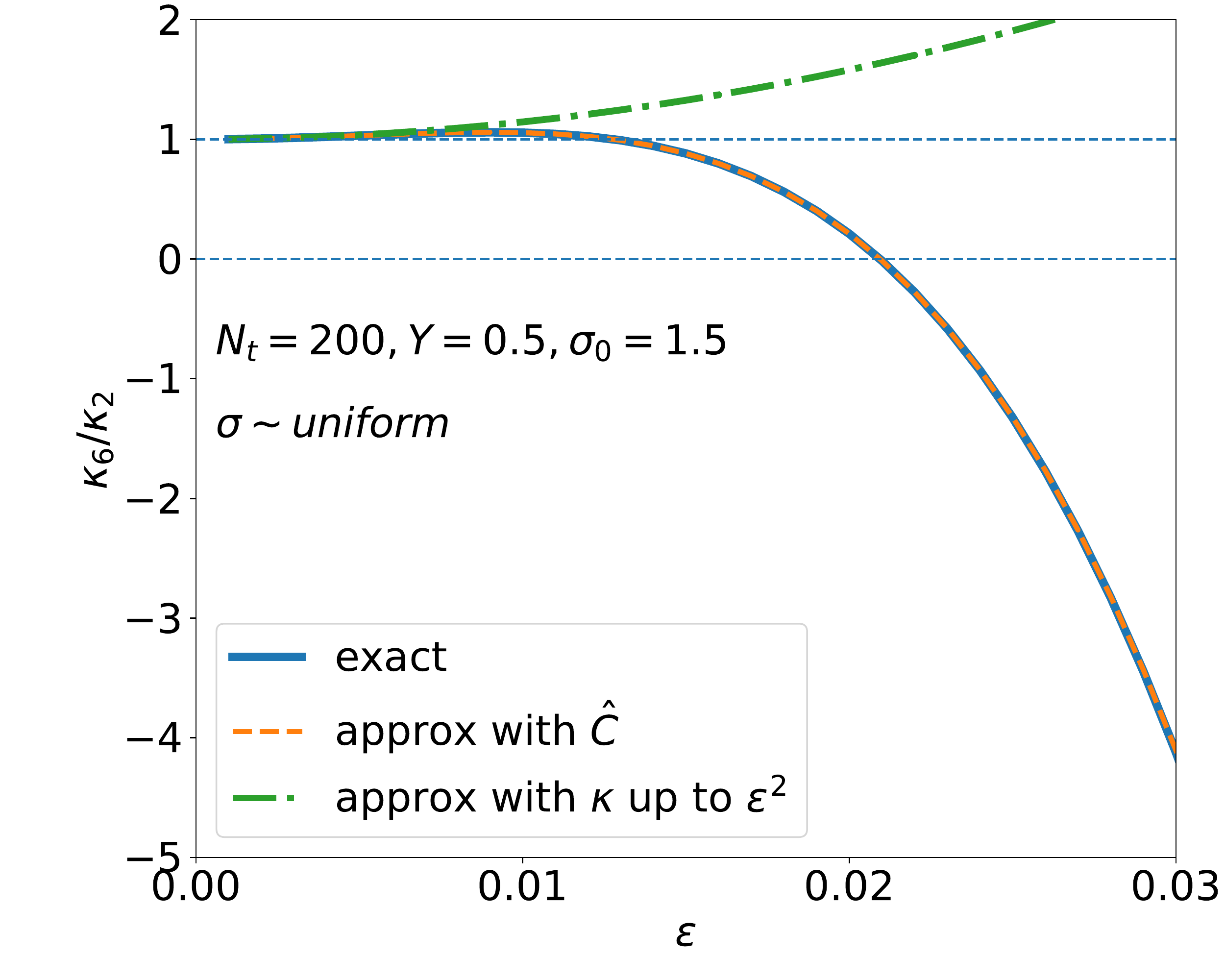}
\caption{$\kappa_5/\kappa_2$ and $\kappa_6/\kappa_2$ for $\msg$ following the uniform distribution. The left-hand side plots show a wider $\meps$ range [0, 0.1] whereas the right-hand side plots show details at small $\meps \in [0, 0.03]$. \label{fig:cumulant-sig-unif56}}
\end{center}
\end{figure}

\begin{figure}[H]
\begin{center}
    \includegraphics[width=0.46\textwidth]{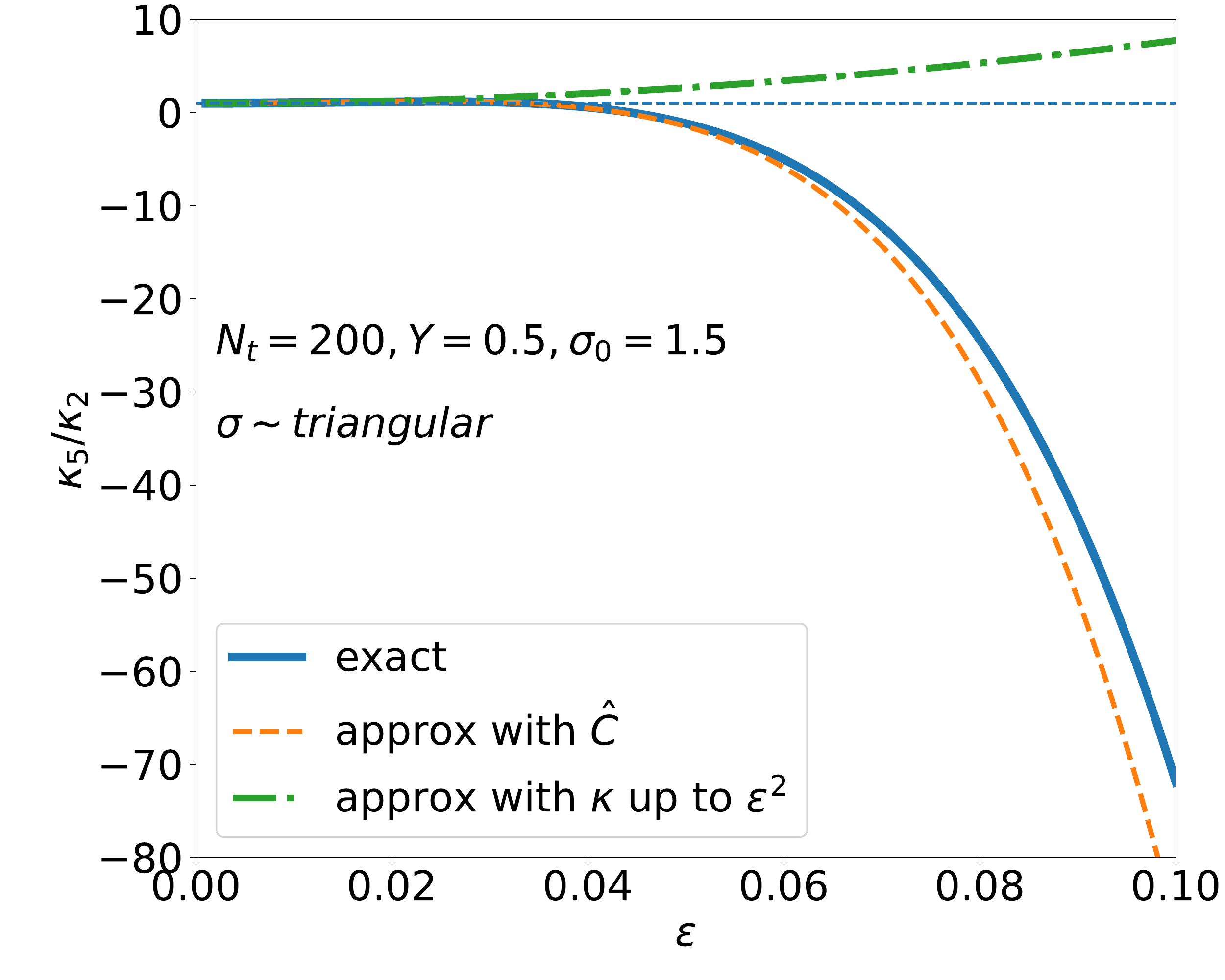}
    \includegraphics[width=0.46\textwidth]{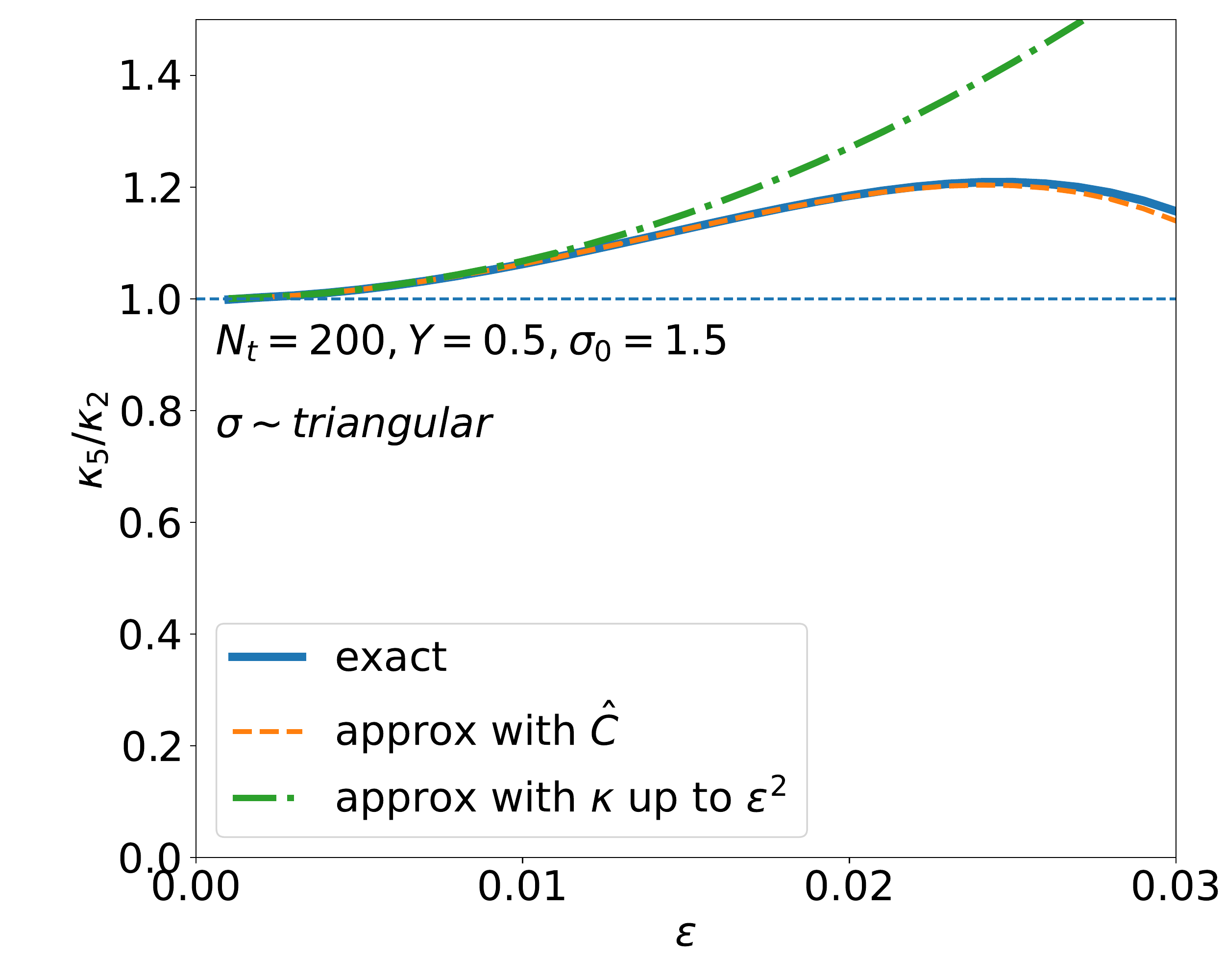}
    \includegraphics[width=0.46\textwidth]{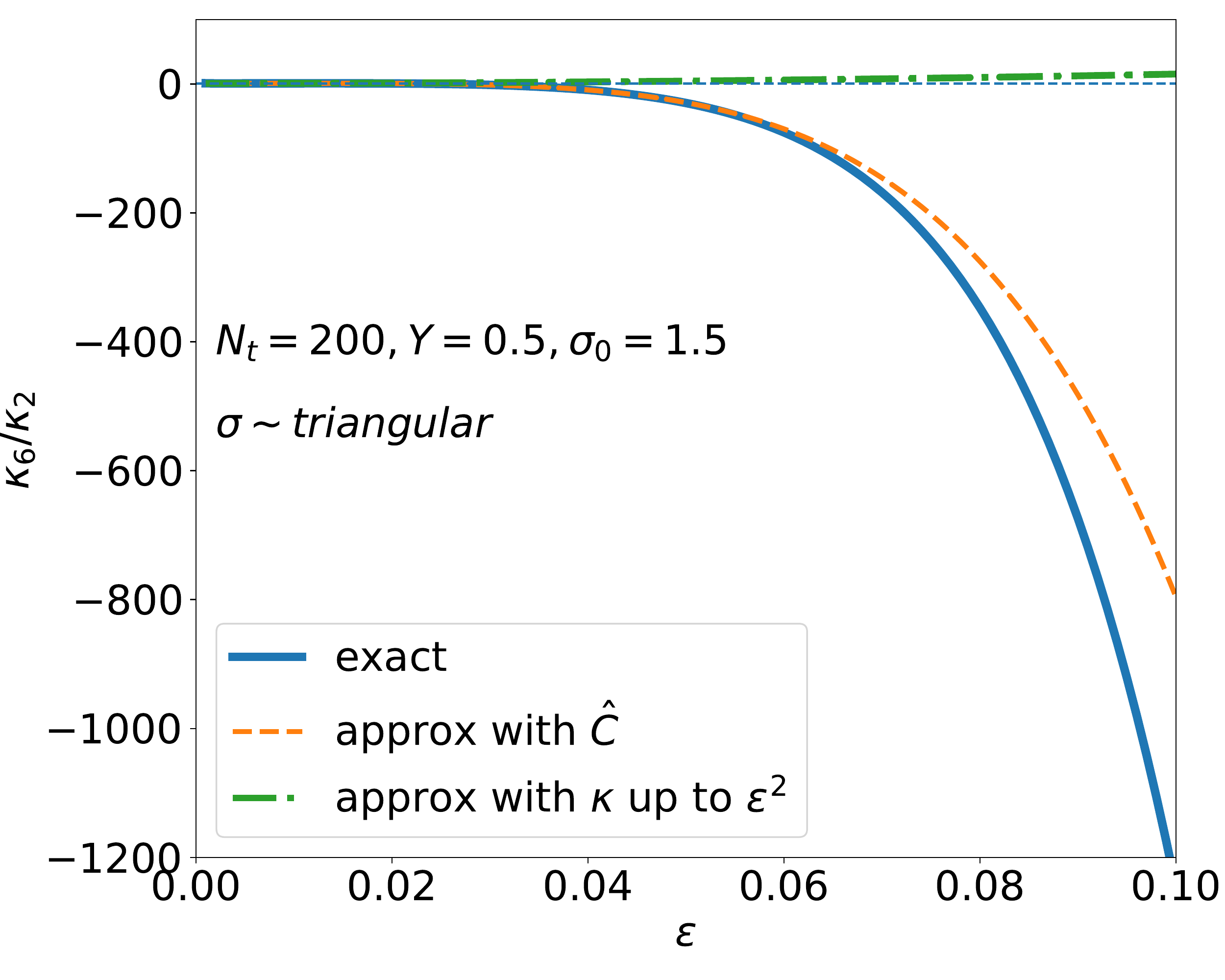}
    \includegraphics[width=0.46\textwidth]{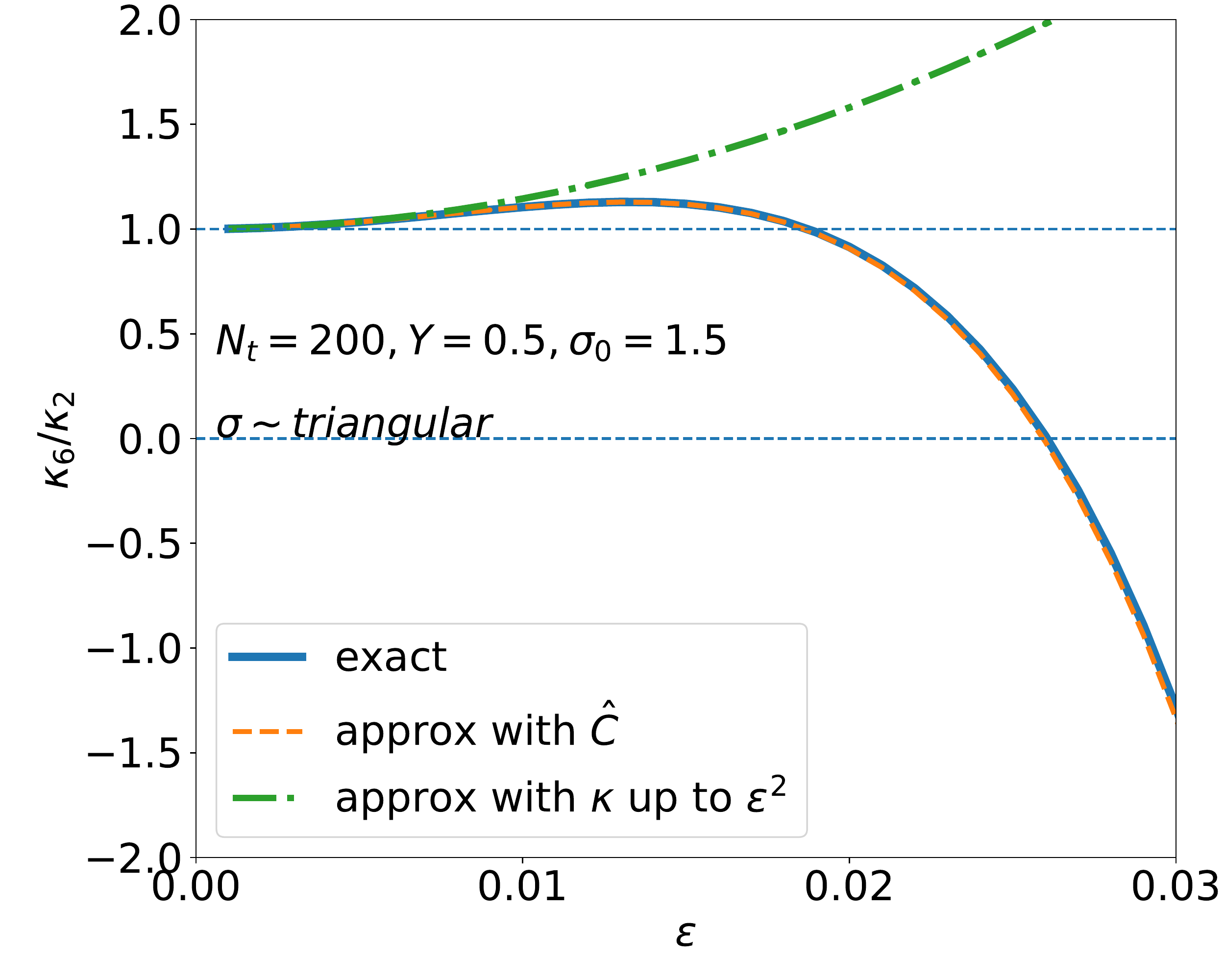}
\caption{Same as Fig. \ref{fig:cumulant-sig-unif56} but for $\msg$ following the triangular distribution. \label{fig:cumulant-sig-trian56}}
\end{center}
\end{figure}

\begin{figure}[H]
\begin{center}
    \includegraphics[width=0.46\textwidth]{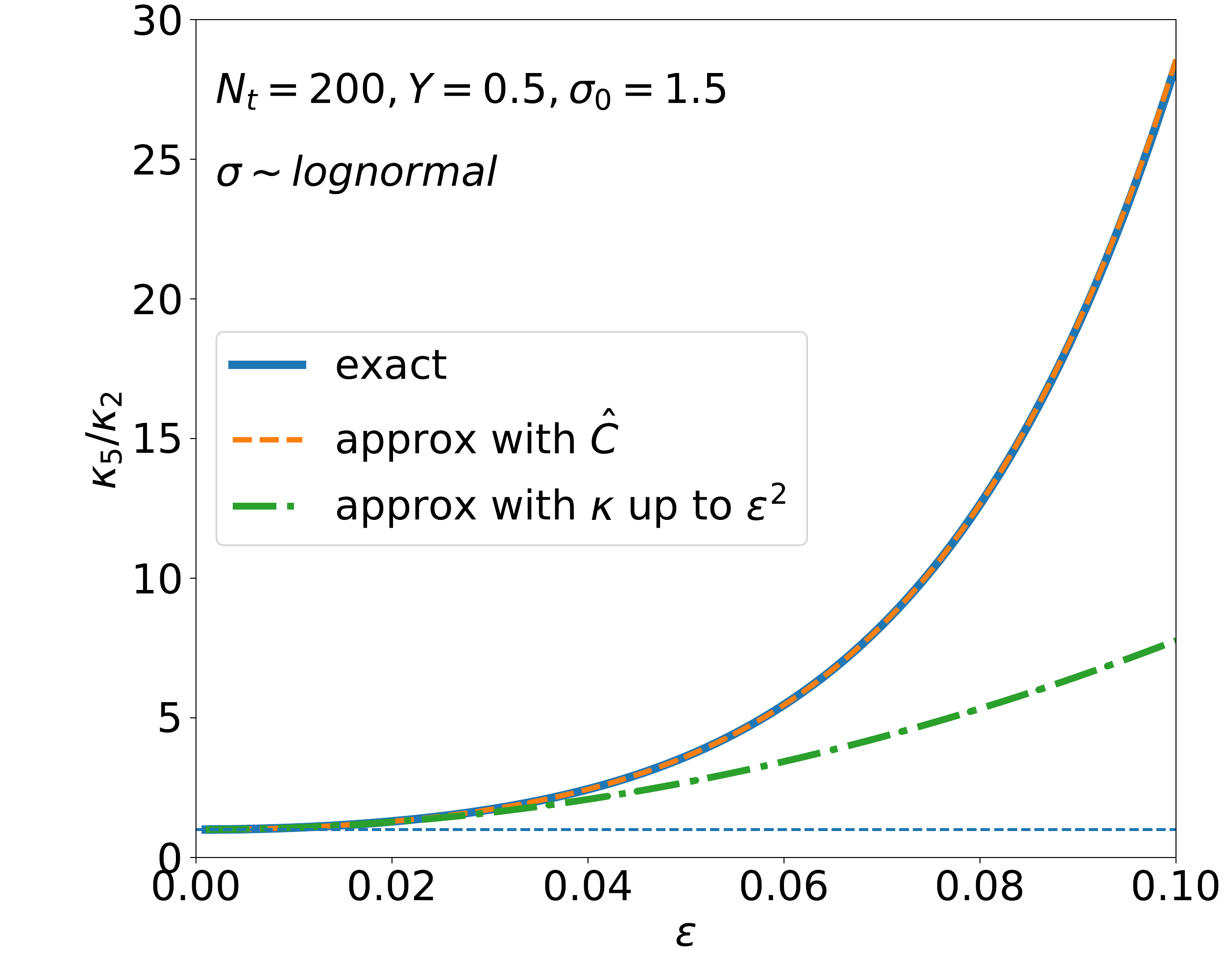}
    \includegraphics[width=0.46\textwidth]{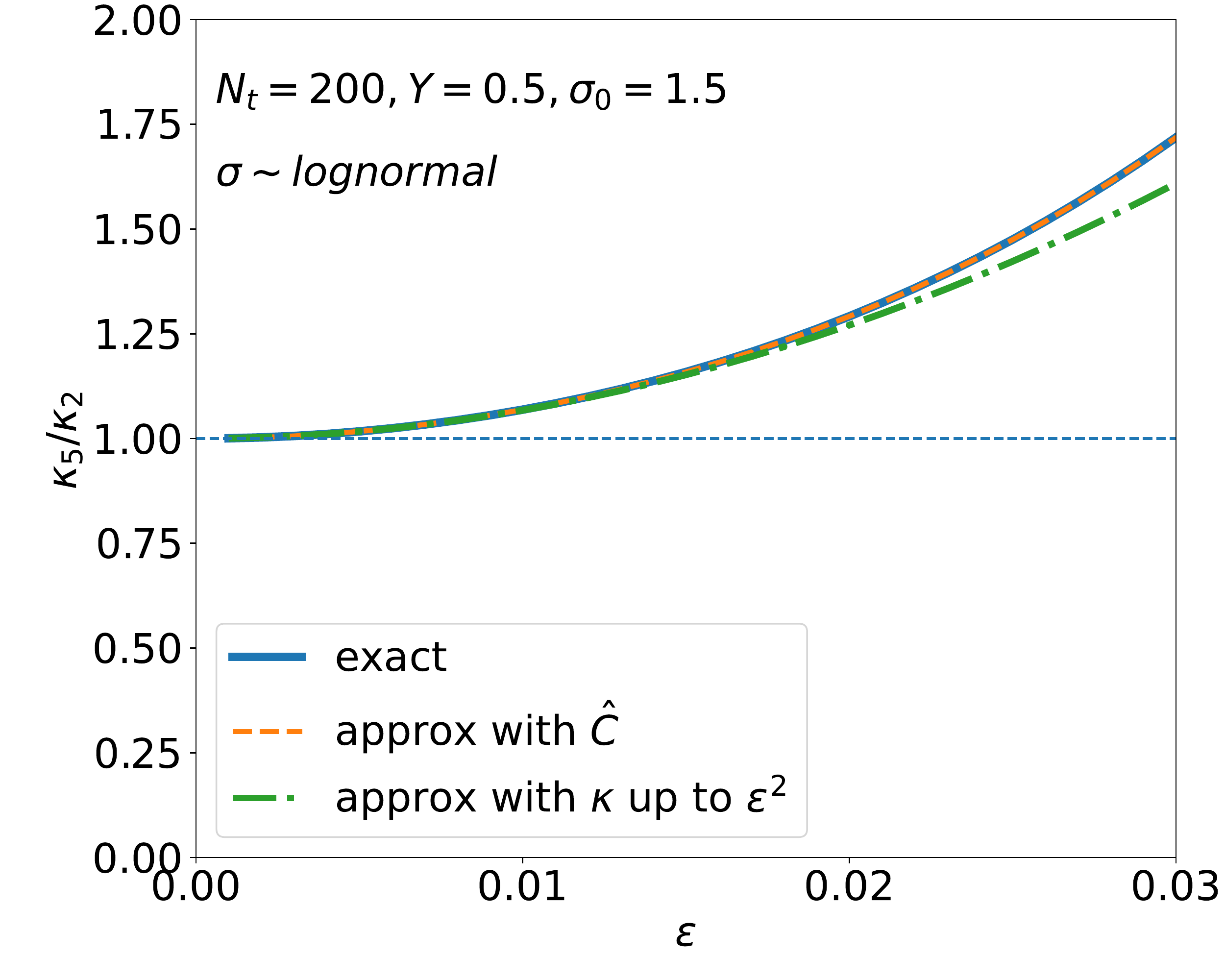}
    \includegraphics[width=0.46\textwidth]{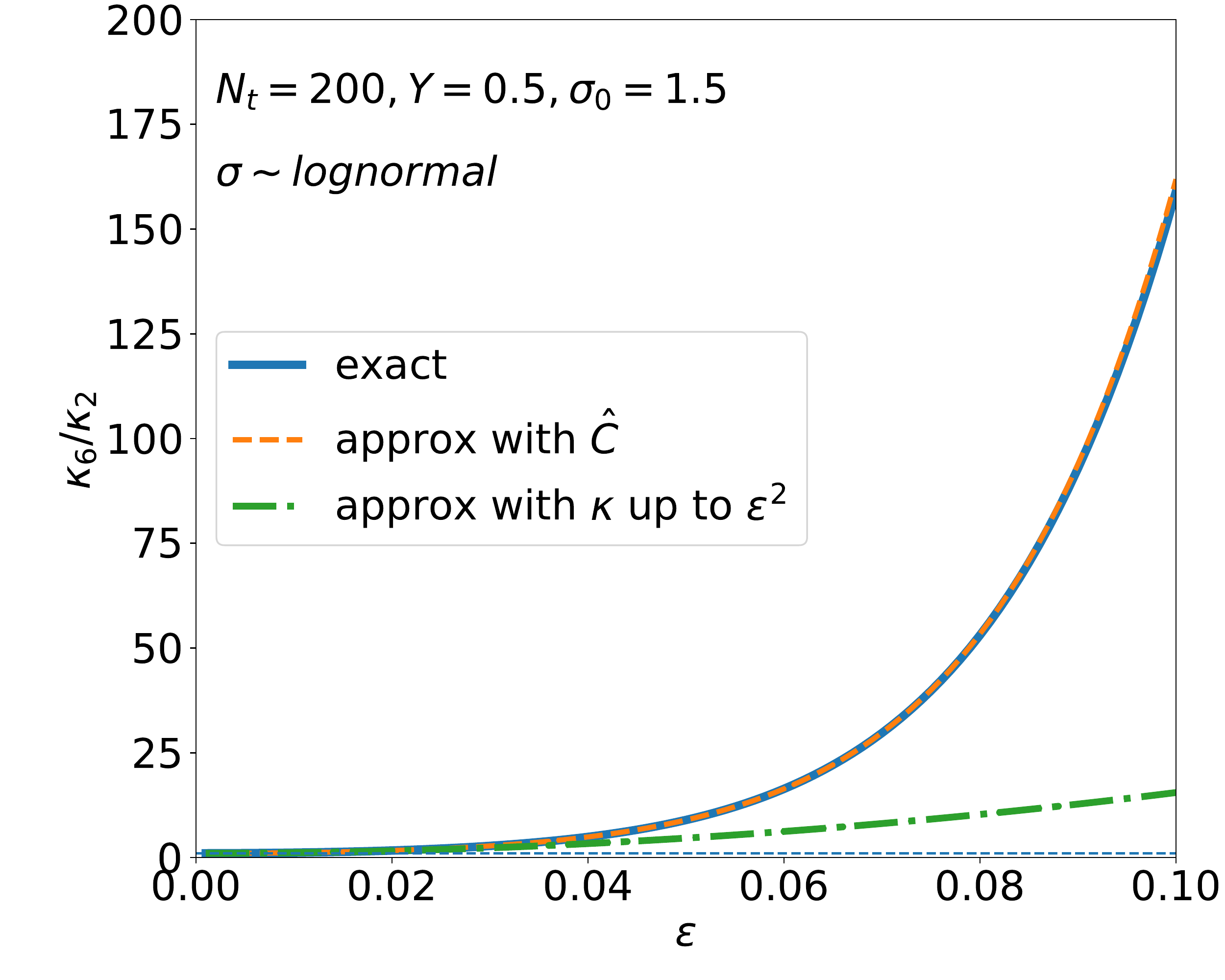}
    \includegraphics[width=0.46\textwidth]{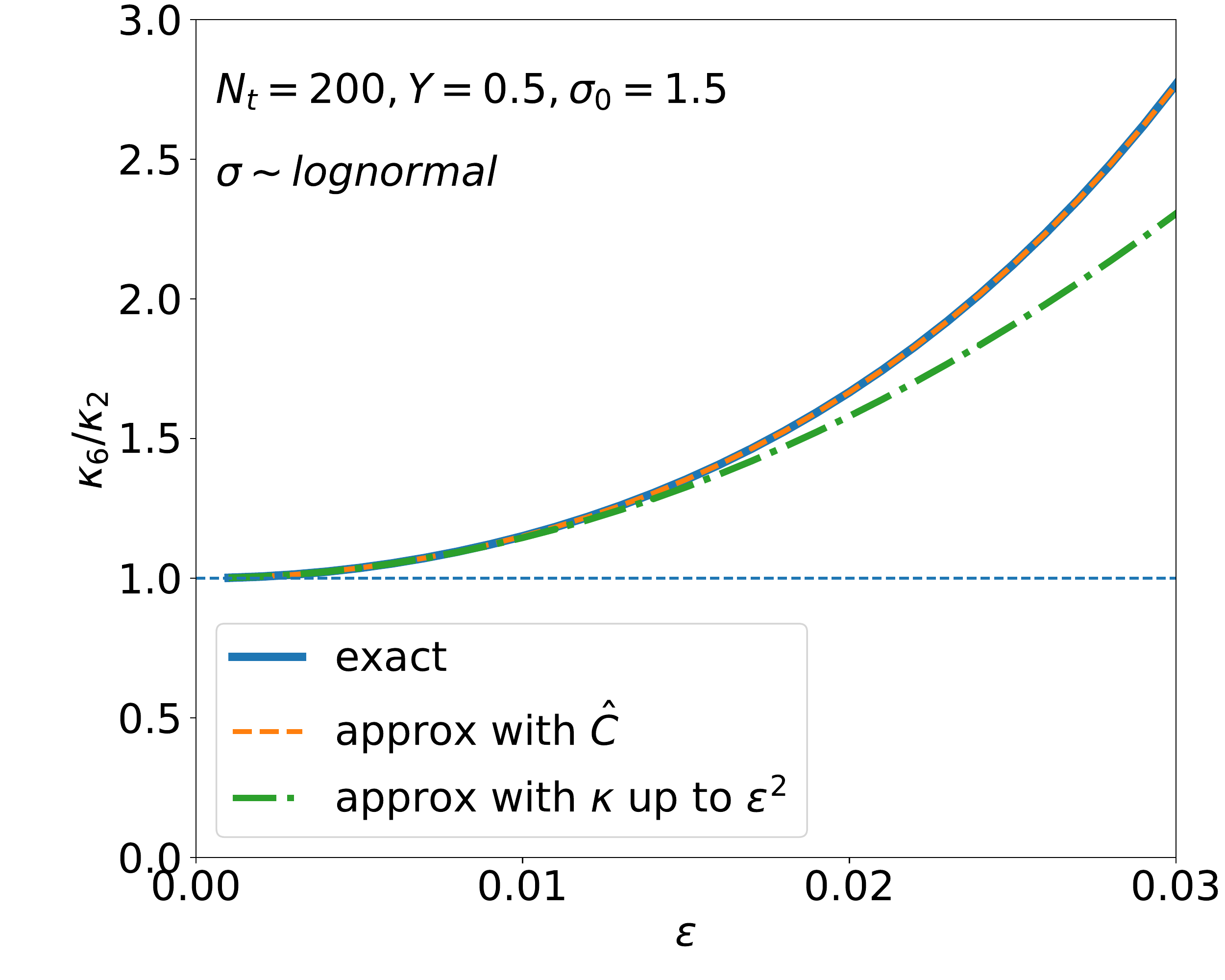}
\caption{Same as Fig. \ref{fig:cumulant-sig-unif56} but for $\msg$ following the lognormal distribution. \label{fig:cumulant-sig-logn56}}
\end{center}
\end{figure}

The cumulant ratios, $\kappa_5/\kappa_2$ and $\kappa_6/\kappa_2$, with $\msg$ following the uniform, triangular, and lognormal distributions are presented in Figs. \ref{fig:cumulant-sig-unif56}, \ref{fig:cumulant-sig-trian56}, and \ref{fig:cumulant-sig-logn56}.

\begin{figure}[H]
\begin{center}
    \includegraphics[width=0.46\textwidth]{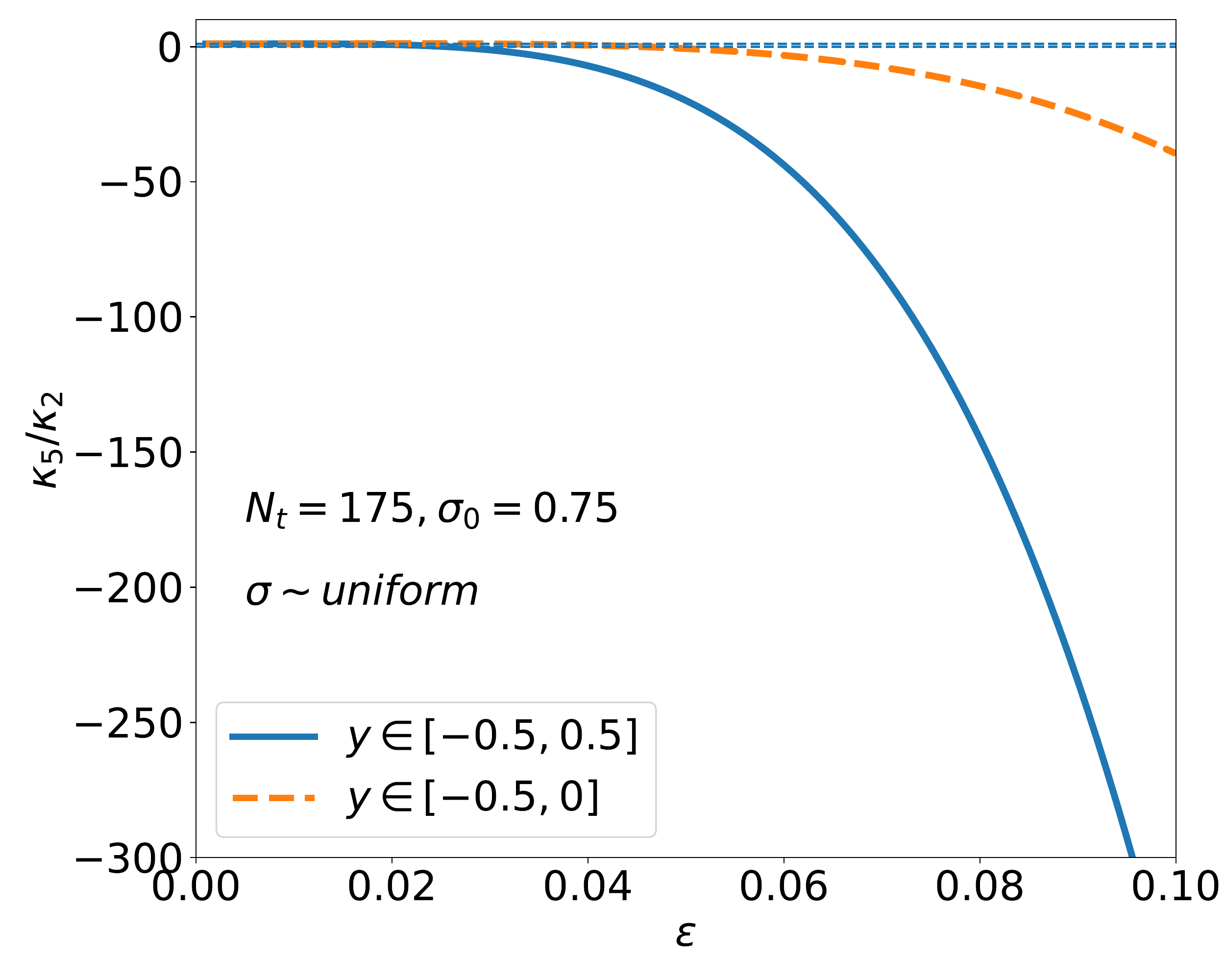}
    \includegraphics[width=0.46\textwidth]{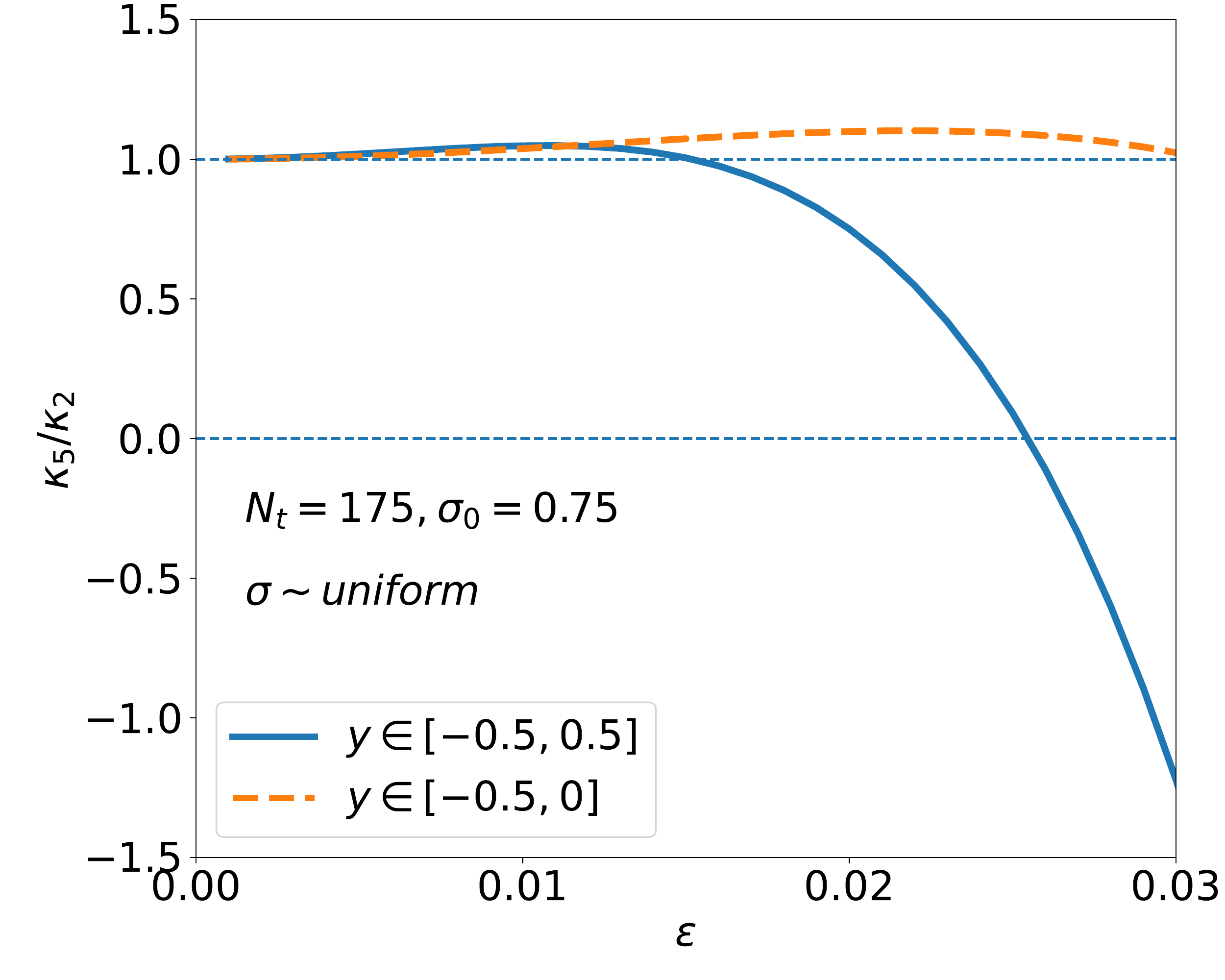}
    \includegraphics[width=0.46\textwidth]{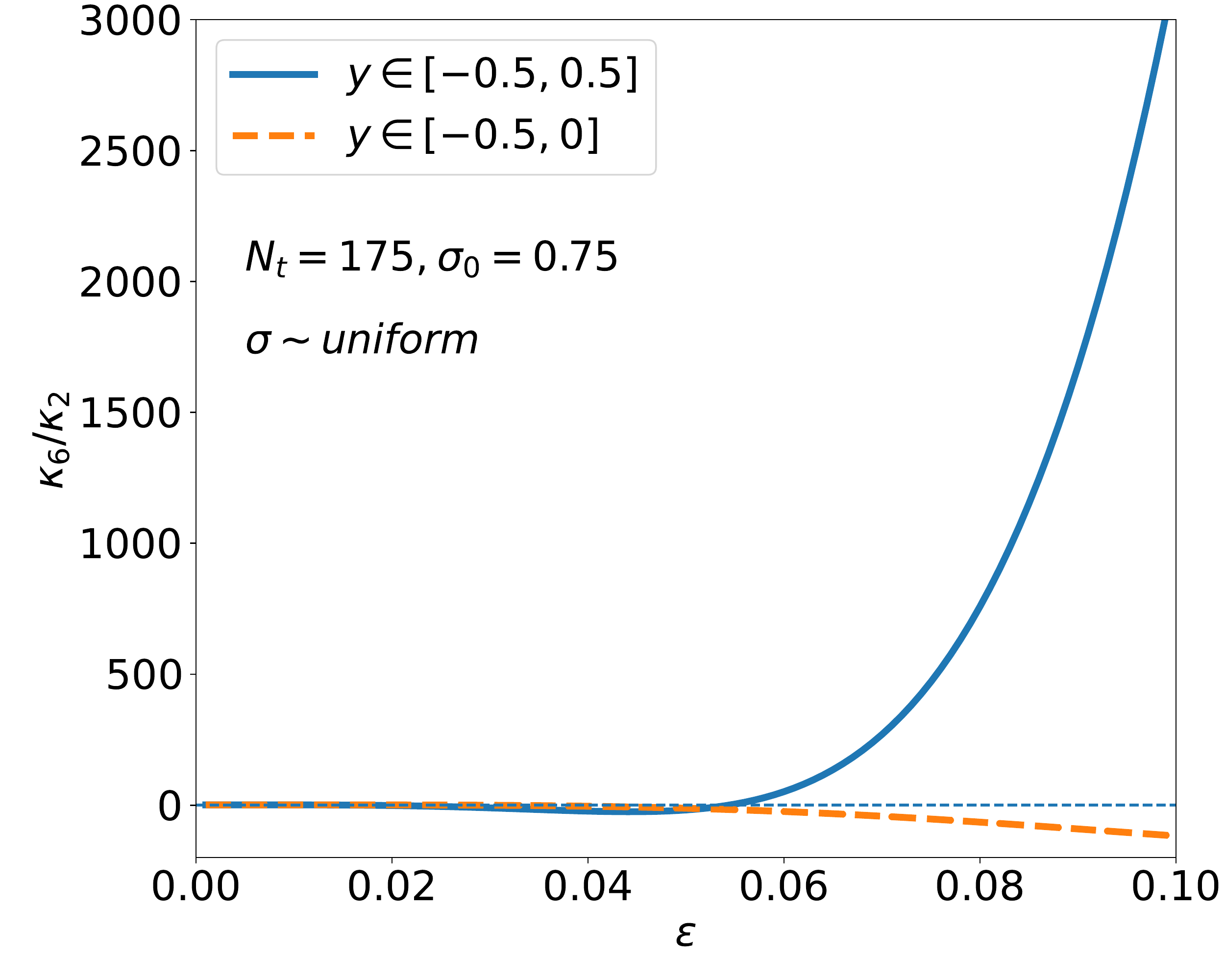}
    \includegraphics[width=0.46\textwidth]{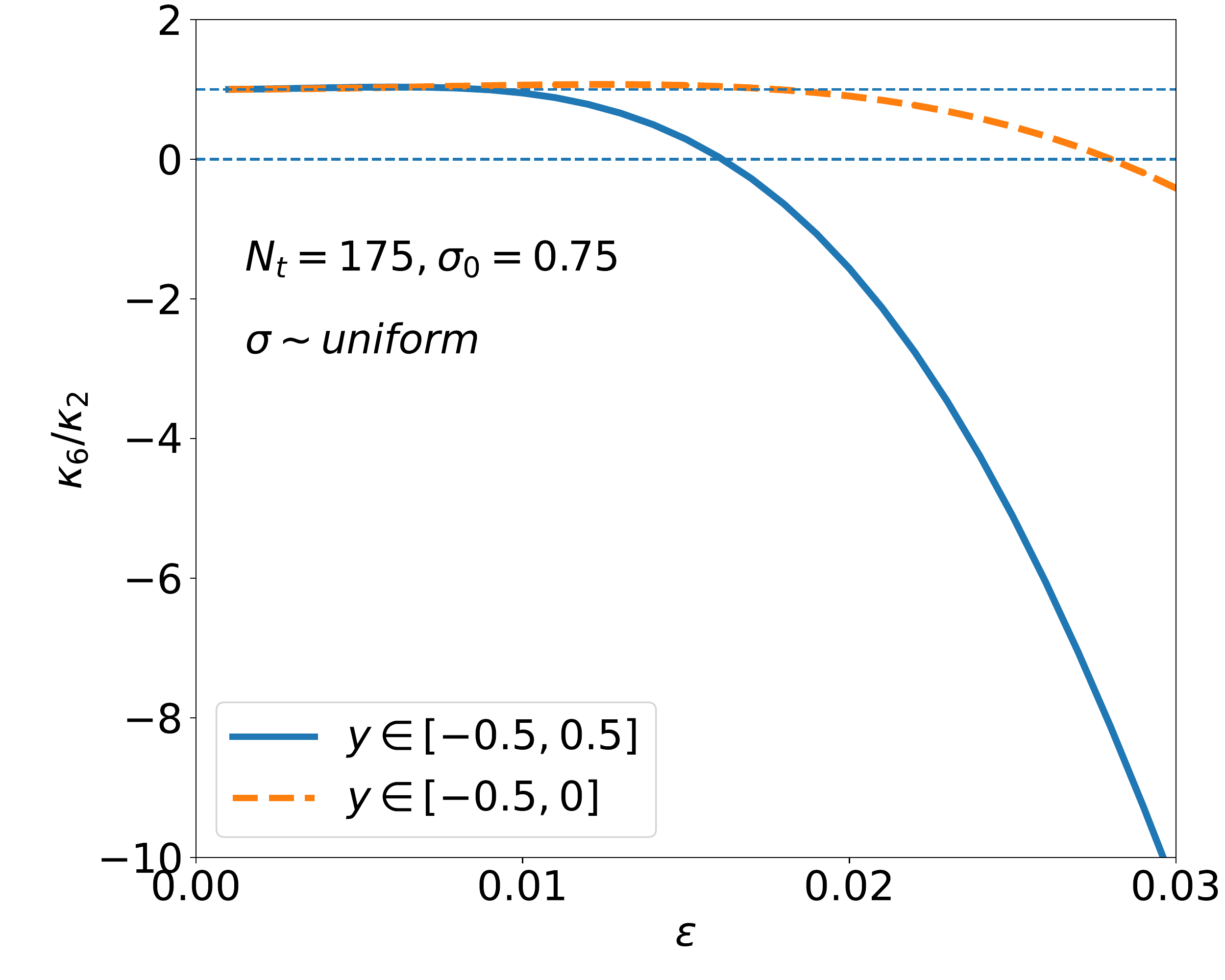}
\caption{$\kappa_5/\kappa_2$ and $\kappa_6/\kappa_2$ for $\msg$ following the uniform distribution. The left hand side plots show wider $\meps$ range [0, 0.1] whereas the right hand side plots show details at small $\meps \in [0, 0.03]$. \label{fig:cumulant-sig-unif56-yrange}}
\end{center}
\end{figure}

\begin{figure}[H]
\begin{center}
    \includegraphics[width=0.46\textwidth]{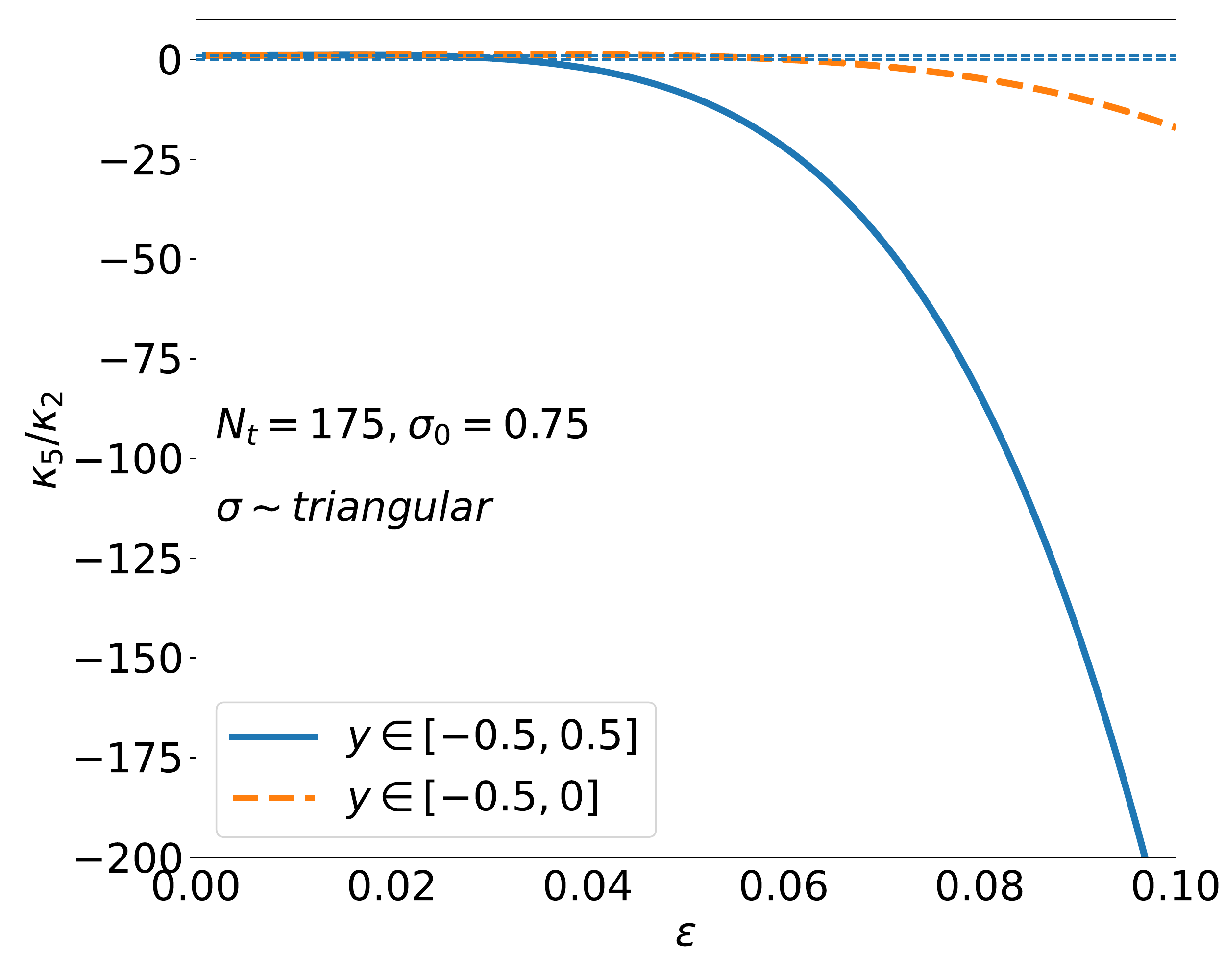}
    \includegraphics[width=0.46\textwidth]{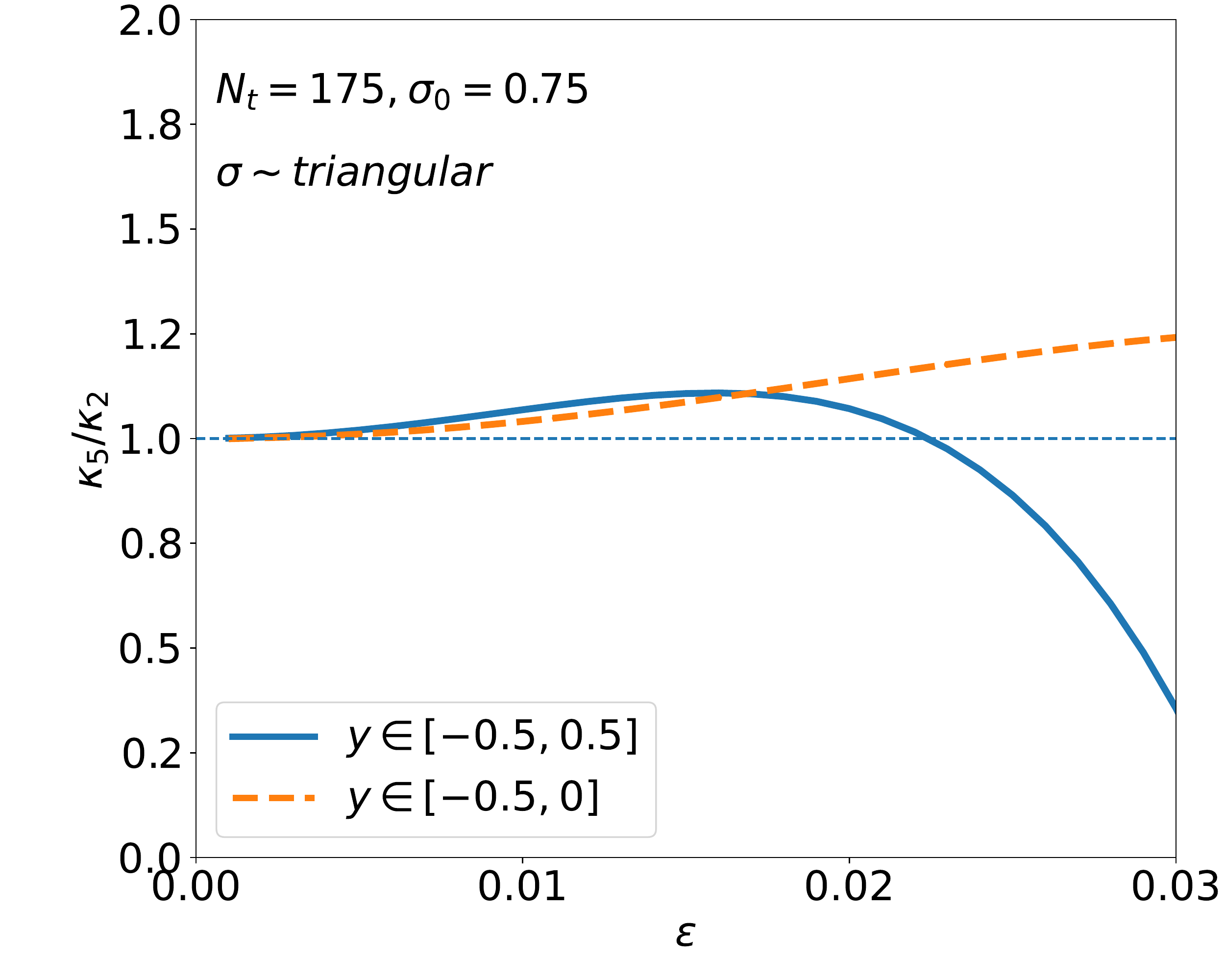}
    \includegraphics[width=0.46\textwidth]{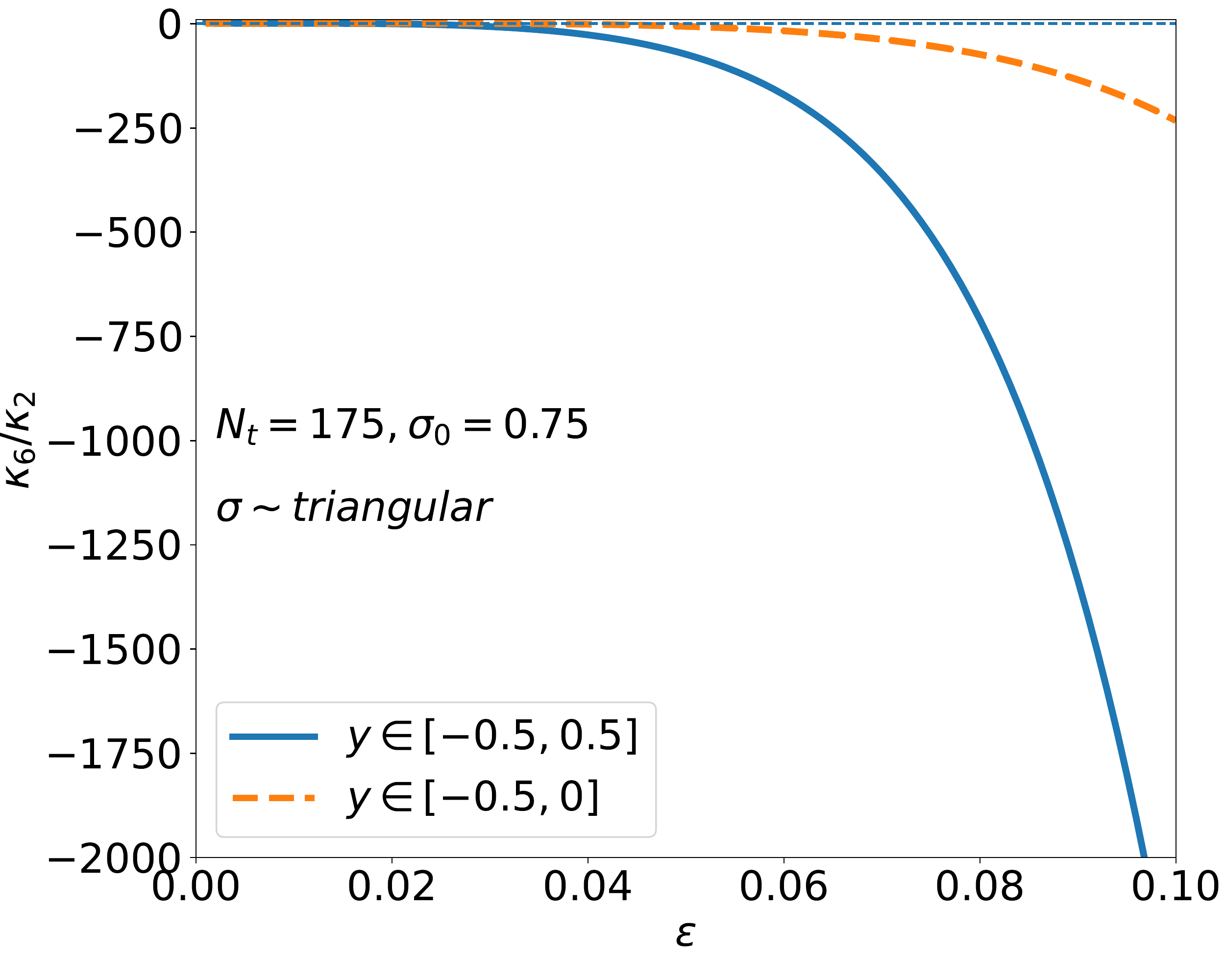}
    \includegraphics[width=0.46\textwidth]{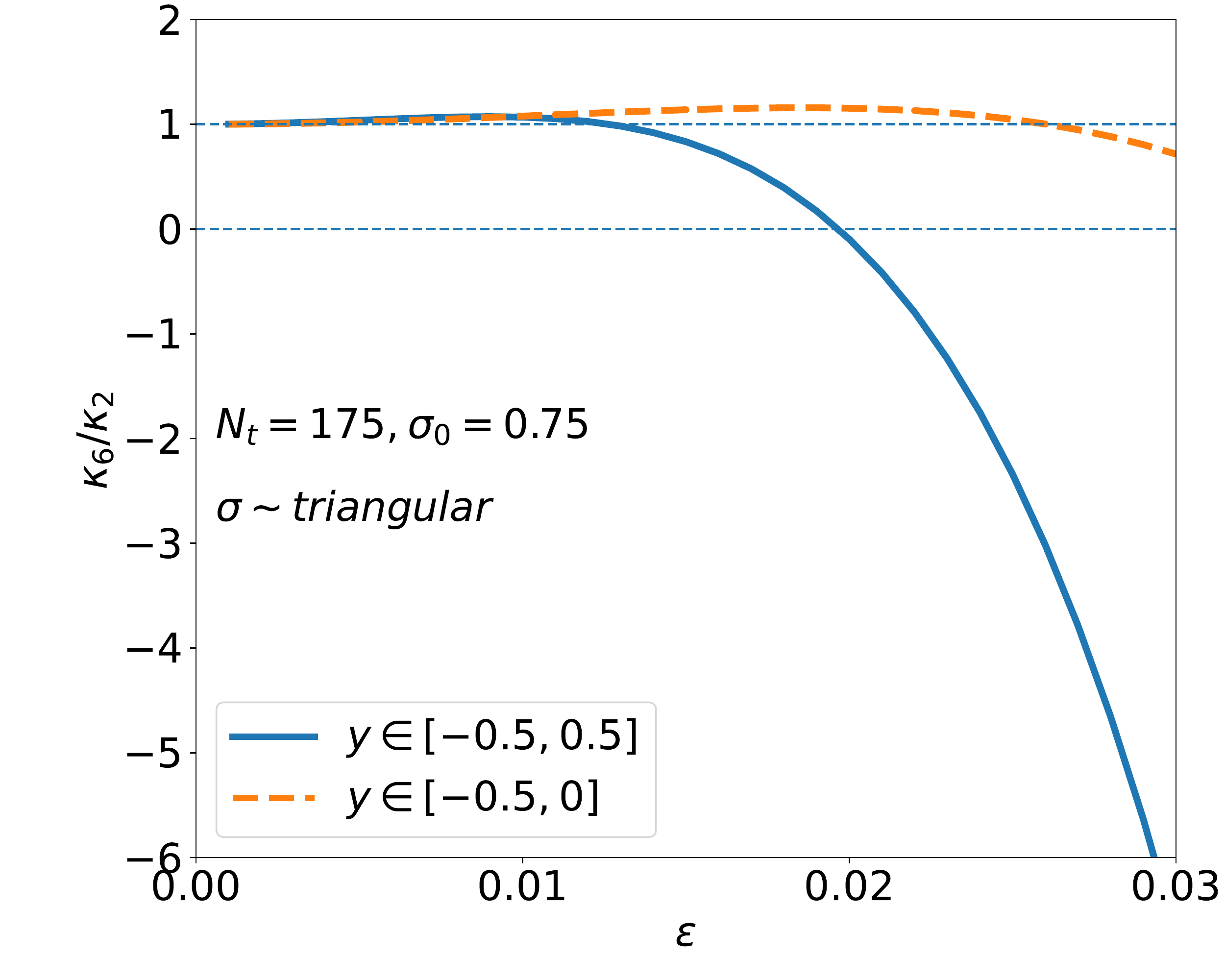}
\caption{Same as Fig. \ref{fig:cumulant-sig-unif56-yrange} but for $\msg$ following the triangular distribution. \label{fig:cumulant-sig-trian56-yrange}}
\end{center}
\end{figure}

\begin{figure}[H]
\begin{center}
    \includegraphics[width=0.46\textwidth]{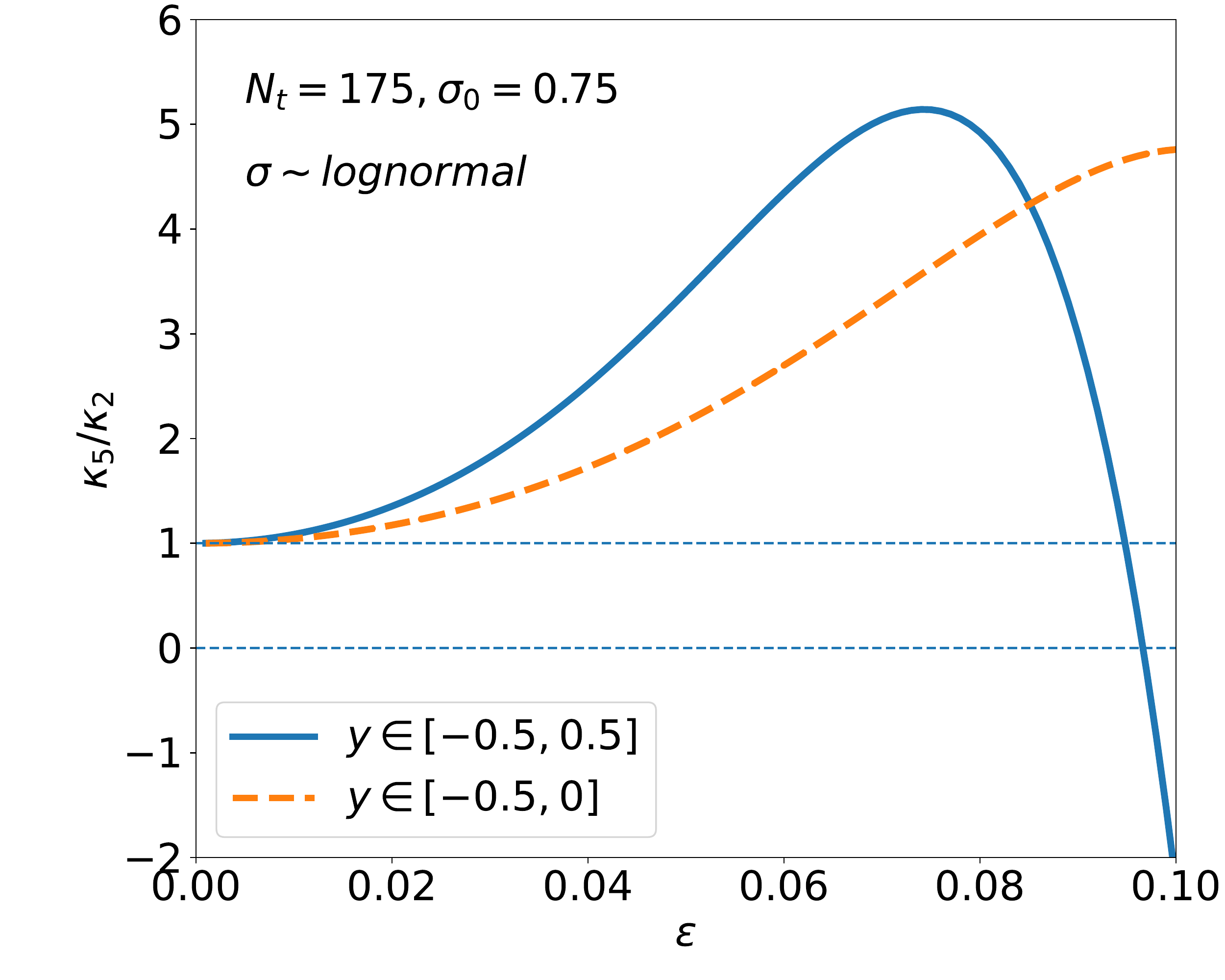}
    \includegraphics[width=0.46\textwidth]{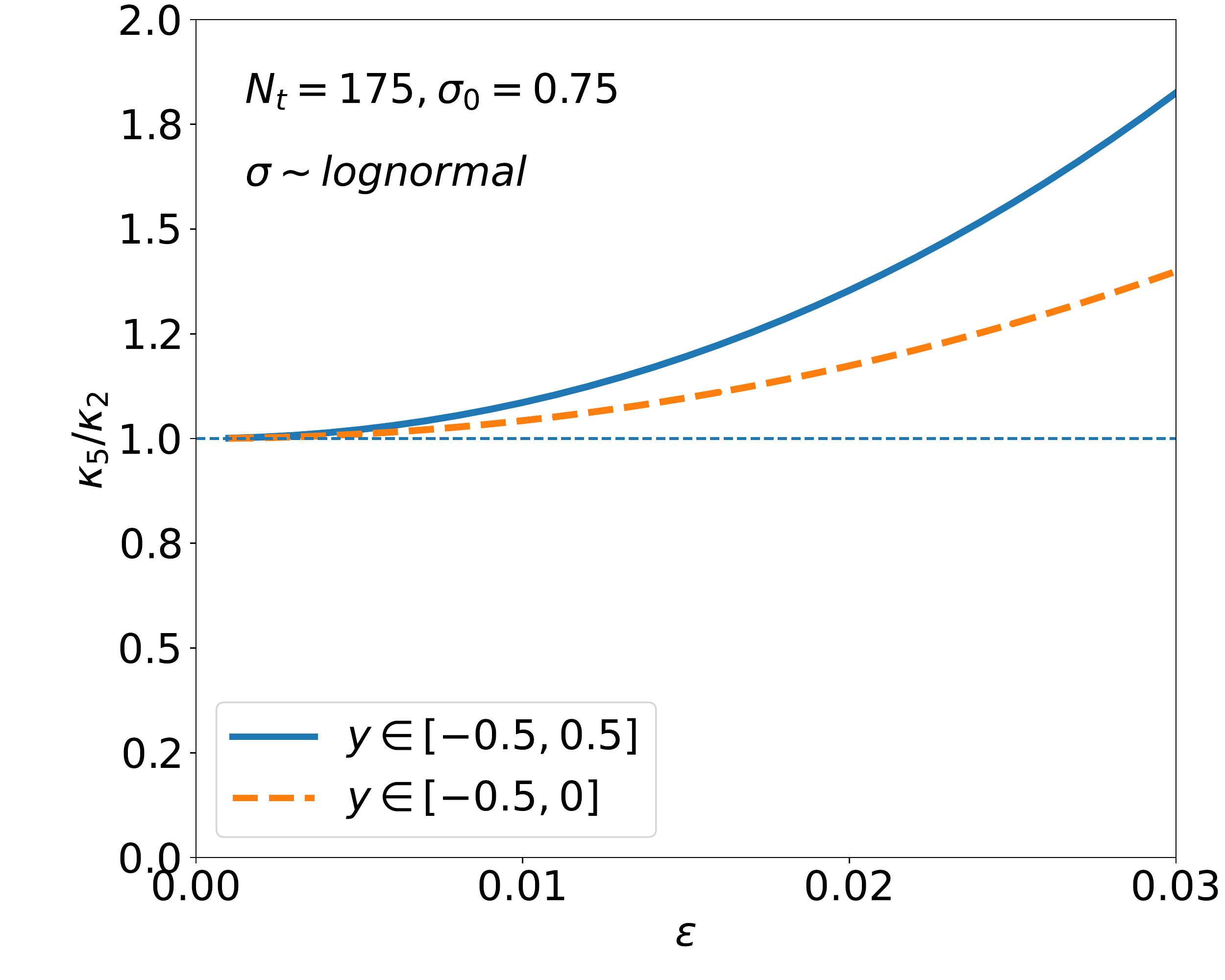}
    \includegraphics[width=0.46\textwidth]{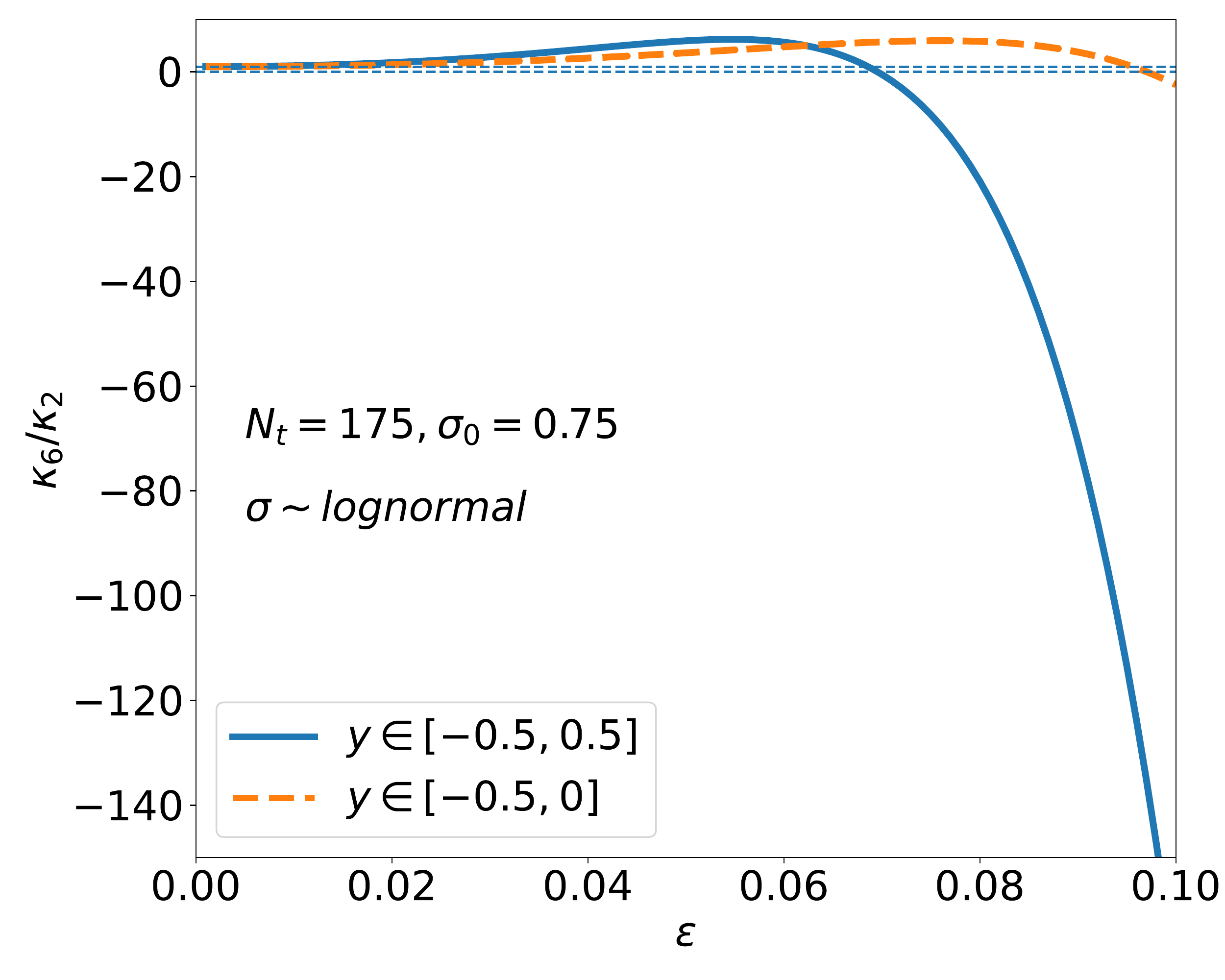}
    \includegraphics[width=0.46\textwidth]{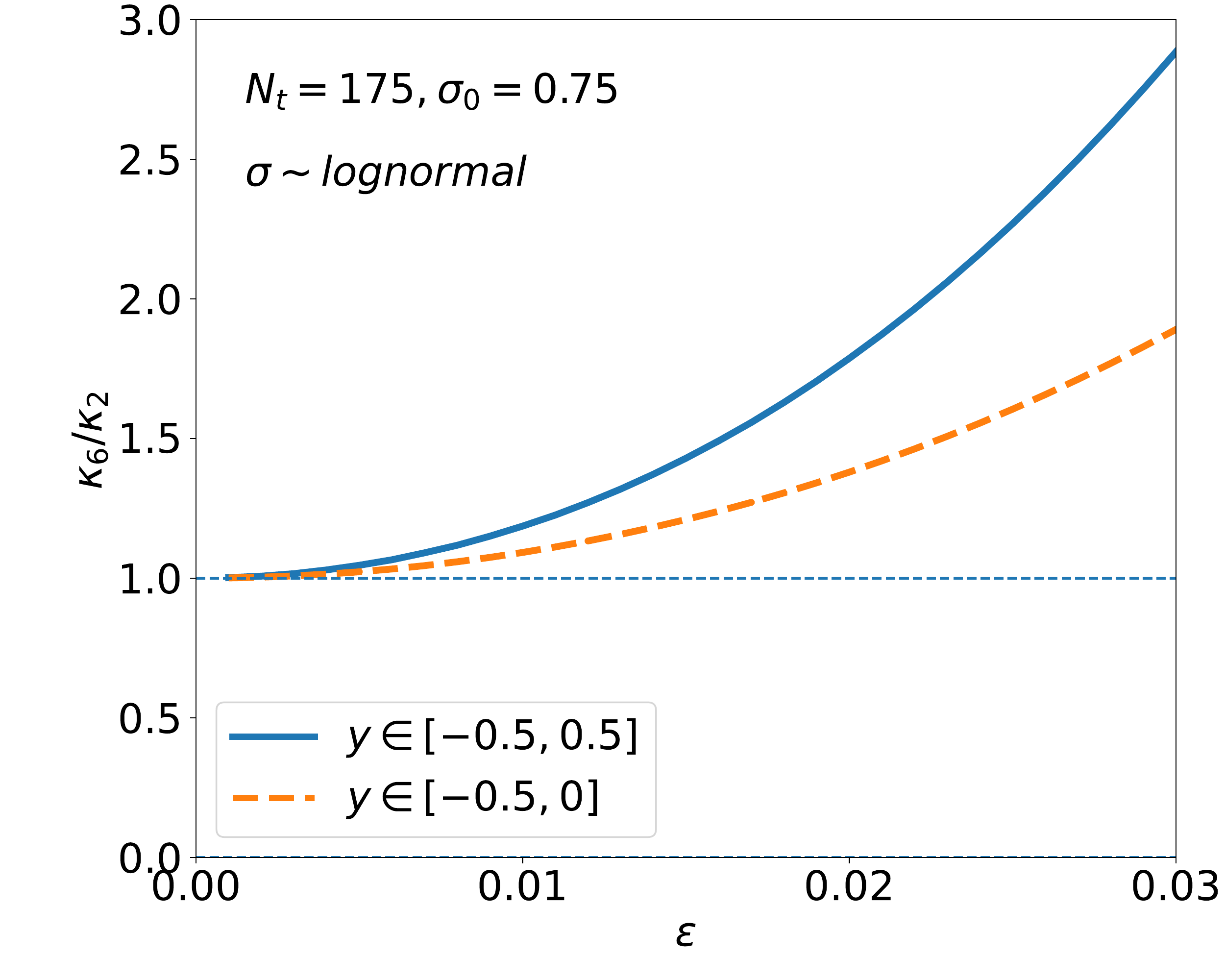}
\caption{Same as Fig. \ref{fig:cumulant-sig-unif56-yrange} but for $\msg$ following the lognormal distribution. \label{fig:cumulant-sig-logn56-yrange}}
\end{center}
\end{figure}

In Figs. \ref{fig:cumulant-sig-unif56-yrange}, \ref{fig:cumulant-sig-trian56-yrange}, and \ref{fig:cumulant-sig-logn56-yrange}, we compare the cumulant ratios, $\kappa_5/\kappa_2$ and $\kappa_6/\kappa_2$, obtained in the symmetric ($y \in [-0.5, 0.5]$) and asymmetric ($y \in [-0.5, 0]$) rapidity interval for all three analytically studied $\msg$ distributions with $N_t$ and $\msg_0$ corresponding to the STAR measurements at 3~GeV.


\bibliography{fluct_width_v4}

\end{document}